\documentclass[twocolumn]{aastex7}

\usepackage{comment}
\usepackage{natbib} 
\usepackage{amsmath}
\usepackage[mathlines]{lineno}

\newcommand{\kms}{km s$^{-1}$}

\newcommand{\hii}{H~{\scriptsize II}}
\newcommand{\cii}{C~{\scriptsize II}}

\newcommand{\degree}{$^{\circ}$}

\usepackage{graphicx} 
\usepackage{multirow}
\usepackage{hyperref}

\begin{document}

\title{The Sagittarius C Complex in the Mid-Infrared with SOFIA/FORCAST}
\shorttitle{MIR View of Sagittarius C}
\shortauthors{Roy J. Zhao et al.}

\makeatletter
\def\@fnsymbol#1{%
  \ensuremath{%
    \ifcase#1\or
      \dagger\or
      \ddagger\or
      \mathsection\or
      \mathparagraph\or
      *\or
      \dagger\dagger\or
      \ddagger\ddagger\or
      \mathsection\mathsection\or
      \mathparagraph\mathparagraph
    \else
      \@ctrerr
    \fi
  }%
}
\makeatother


\author[orcid=[0009-0001-9716-4188,sname='Zhao']{Roy J. Zhao}
\altaffiliation{E-mail: \hyperlink{mailto:rzhaolx@stanford.edu}{rzhaolx@stanford.edu}. Now at Kavli Institute for Particle Astrophysics \& Cosmology (KIPAC) and Department of Physics at Stanford University, Stanford, CA 94305, USA}
\affiliation{Kavli Institute for Cosmological Physics, The University of Chicago, 5640 S Ellis Ave., Chicago, IL 60637, USA}
\affiliation{Department of Physics, The University of Chicago, 5720 S Ellis Ave., Chicago, IL 60637, USA}
\affiliation{NSF-Simons AI Institute for the Sky (SkAI), 172 E. Chestnut St., Chicago, IL 60611, USA}
\email[]{rzhaolx@stanford.edu} 

\author[orcid=[[0000-0002-6753-2066,sname='Morris']{Mark R. Morris}
\affiliation{Department of Physics \& Astronomy, UCLA, 475 Portola Pl., Los Angeles, CA 90095, USA}
\email[]{morris@astro.ucla.edu} 

\author[orcid=[0000-0001-9315-8437,sname='Hankins']{Matthew J. Hankins}
\affiliation{Arkansas Tech University, 215 West O Street, Russellville, AR 72801, USA}
\email[]{mhankins1@atu.edu} 

\author[orcid=[0009-0001-9323-971X,sname='Cotera']{Angela S. Cotera}
\affiliation{SETI Institute, 189 Bernardo Ave., Mountain View, CA 94043, USA}
\email[]{ascotera@gmail.com}

\author[orcid=[0000-0001-8095-4610,sname='Simpson']{Janet P. Simpson}
\affiliation{SETI Institute, 189 Bernardo Ave., Mountain View, CA 94043, USA}
\email[]{janet.p.simpson@gmail.com} 

\begin{abstract}
We present an analysis of high-resolution mid-infrared observations at 25 and 37 \micron{} of the Sagittarius C Complex (Sgr C) in the Central Molecular Zone, based on data from the SOFIA/FORCAST Galactic Center Legacy Survey. 
Enabled by the high bright-source limit of the FORCAST instrument, we perform a pixel-level dust temperature and optical depth map with a focus on the Sgr C \hii{} region, which has an average dust temperature of $61$ K and average 37 \micron{} optical depth of $0.05$. 
The Sgr C \hii{} region contains several high-density dust emission ridges, with lengths of up to several parsecs. 
Noting prior evidence for nonthermal radio emission from these ridges, we postulate that there is an enhancement of relativistic electrons within them, possibly attributable to diffusive shock acceleration induced by the impact of wind from a known nearby Wolf-Rayet (WR) star.
We examined the heating effect of the WR star by calculating its heating profile, finding an agreement between the profiles and our dust temperature map assuming smaller grains ($\sim 0.01$\micron{}). 
We perform a spectral energy distribution modelling of the \hii{} region and found an integrated MIR luminosity of $(1.40\pm0.19)\times10^{6} L_\odot$, which implies that presently unidentified massive stars must be present in the \hii{} region in addition to the WR star.
We finally survey adjacent regions, such as a mid-infrared/radio source denoted ``Source C" and the G359.43+0.02 young stellar object cluster near the northern end of the prominent Sgr C non-thermal radio filament. 
\end{abstract}


\section{Introduction}
The central half kiloparsec of the Milky Way disk is known as the Central Molecular Zone (CMZ) for its large molecular cloud mass, high cloud densities, strong magnetic fields, and dynamically complex behaviour, but has a star formation rate significantly lower than expected in such extreme environments { \citep[see][and references therein]{Morris1996, Ferriere2007, Sormani2020, Tress2020, Battersby2020, Bryant2021, Henshaw2023, Tress2024}}. Drawing a coherent picture of the astrophysical processes in this region has been a driving objective in the community. As the only major star-forming region at negative Galactic longitudes in the CMZ \citep{Yusef-Zadeh2009, Lu2019SFR}, located 0.5\degree{} ($\sim$ 75 pc) in projection from the Galaxy's dynamical centre, the Sagittarius C (Sgr C) complex exhibits an intriguing astro-geography, including a prominent \hii{} region, an extended radio non-thermal filament (NTF) over 30 parsecs in length, and a dense, star-forming molecular cloud in the vicinity. As such, Sgr C is a valuable laboratory to study the interactions between these astrophysical components, therefore contributing to our understanding of the CMZ as a whole. 

Historically, Sgr C has received less attention than its more luminous and star-bursting counterpart, Sgr B2, at the opposite end of the CMZ. Very recently, however, a series of studies of Sgr C has been conducted with data from the Stratospheric Observatory for Infrared Astronomy (SOFIA) and the \textit{James Webb Space Telescope} {(\it JWST)} \citep[e.g.,][]{Bally2024, Crowe2024, DeBuizer2025, Riquelme2026, Zhao2025}. Given the current momentum of the community in studying this region, we aim here to further characterize the stellar population and the interstellar medium (ISM) properties within this region in the MIR regime. 

In the past decade, many studies have revealed high star formation activity within the Sgr C complex at large. Most notably, in the southeast side\footnote{Cardinal directions used in this paper assume a Galactic coordinate system.} of the prominent Sgr C \hii{} region (G359.43-0.09; first reported by \citealt{Downes1966}), \citet{Kendrew2013} reported two massive and compact protostellar cores, G359.44-0.10, signified by their excess 4.5 \micron{} emission (so-called extended green objects, or EGOs; \citealt{Cyganowski2008}). {Far-IR observations reveal that these cores reside in a particularly high density ``tip" of a dense molecular cloud \citep{Hatchfield2020, Zhao2025}, which follows the famous 100 parsec dust ring \citep{Molinari2011}. \citet{Lu2019SFR} suggested that the EGO is likely interacting with the \hii{} region, resulting in the formation of protostellar objects. In addition, \cite{Kendrew2013} found no evidence for star formation outside of the EGO. Therefore, this cloud is mostly quiescent unlike its mini-starbursting sister Sgr B2 \citep{Yusef-Zadeh2009,Belloche2013, Ginsburg2018}.} These cores were very recently studied by \citet{Crowe2024} using JWST-NIRCam imaging in conjunction with data from the Atacama Large Millimeter/submillimeter Array (ALMA), finding that these cores exhibit high bolometric luminosity ($\sim 10^5 L_\odot$) and high cloud mass ($\sim 100 M_\odot$). A series of spectroscopic and radio studies also found various manifestations of star formation within and around the EGO, including ${\rm CH_3OH}$ and ${\rm H_2O}$ masers with radio counterparts, protostellar clouds undergoing thermal fragmentation, and outflows from these protostars \citep{Lu2019, Lu2020, Lu2021, Rickert2019}.

To the north of the Sgr C \hii{} region, the spectral energy distribution (SED) analysis performed by \citet{Yusef-Zadeh2009} revealed a cluster of YSOs 0.1 degree north of the Sgr C \hii{} region, denoted as the G359.43+0.02 cluster. These YSOs are found to be in an earlier evolutionary stage than those in Sgr B2 near the eastern end of the CMZ \citep{Nogueras-Lara2024}. 

Many aspects of the Sgr C \hii{} region, however, still remain relatively poorly characterized, due in part to instrumental limitations. The low saturation limit of the {\it Spitzer}/Multiband Imaging Photometer (MIPS) prevented investigators from studying bright MIR sources, which likely host and are heated by stars, such as the Sgr C \hii{} region and several MIR sources in its vicinity \citep{Carey2009, Yusef-Zadeh2009}. Past efforts to constrain the stellar population with observations at other wavelengths indicated that the Sgr C \hii{} region might contain at least one O4-O6 star, which has driven the formation of its quasi-spherical morphology \citep{Odenwald1984, Liszt1995, Lang2010}. More recently, \citet{Nogueras-Lara2024} reported that there are several $10^5 M_\odot$ of young stars $\sim 20$ Myr old in the immediate environment of the Sgr C \hii{} region. Overall, the {\it Spitzer}/Infrared Spectrograph (IRS) observation reported by \citet{Simpson2018} suggested that the age of stars in the Sgr C region is around $10^{6.6}$ years, consistent with a combination of both current star formation and indications of evolved stars. A recent investigation of the dynamics and morphology of the \hii{} region using JWST-NIRCam to study Br $\alpha$ emission revealed many filamentary structures inside the \hii{} region \citep{Bally2024}. Those authors hypothesized that the ambient magnetic field played a dominant role in the formation of such filaments, and they posited that the filaments likely trace ionization fronts. 

Further motivating a detailed study of the Sgr C MIR sources is the prominent NTF connecting the \hii{} region and the G359.43+0.02 cluster in projection. This NTF, commonly denoted the Sgr C NTF and initially detailed by \citet{Liszt1995}, is one of the most prominent in the Galactic Centre (GC). \citet{Zhao2025} reported signs of interaction between this NTF and the \hii{} region ($b\approx-0.07$\degree{}) and between the NTF and a horizontal radio structure denoted FIR-4 ($b\approx0.01$\degree{}), which is likely associated with the mentioned G359.43+0.02 cluster. Curiously, also near this cluster, this NTF progresses from a single narrow, luminous streak near the \hii{} region to a bundle of diffuse, faint parallel filaments. Thus, studying the MIR sources along the Sgr C NTF may shed light on the physical mechanisms of NTFs within the CMZ in general. 

In this paper, we use the MIR measurement from the Faint Object Infrared Camera for the SOFIA Telescope (FORCAST; \citealt{Herter2012}), which had a greater bright source limit and higher spatial resolution compared to MIPS, albeit with lower sensitivity, to examine in detail these bright MIR regions that were previously saturated in {\it Spitzer}/MIPS. We aim to leverage the inferred dust properties in the MIR-bright regions to study this region. The paper is organized as follows. In Section~\ref{sec:obs}, we outline the observational specifics of the SOFIA/FORCAST Galactic Center Legacy Survey and introduce additional processing on the survey mosaics from previous data releases. Section~\ref{sec:analysis} presents an analysis of the Sgr C complex using FORCAST data, including dust temperature, optical depth, and an examination of individual sources. In Section~\ref{sec:discussion}, we combine the FORCAST data with an ensemble of existing multi-wavelength studies and infer the astrophysical properties of structures within the Sgr C complex, before laying out our conclusions in Section~\ref{sec:conclusion}. {Throughout this paper, we adopt 8.2 kpc as the distance to the GC \citep{GRAVITY2019}. }

\section{Observations}
\label{sec:obs}
{In this work, we use joint observations from two SOFIA programs from two different cycles: the SOFIA/FORCAST Galactic Center Legacy Survey Program carried out in cycle 7 (Plan ID: 07\_0189; PI: M. Hankins), and later observations carried out in cycle 9 (Plan ID: 09\_0216; PI: M. Hankins) designed to complement the Legacy Survey with consistent data quality.} An overview of the Legacy Survey is described in \cite{Hankins2020}, with cycle 9 observations and subsequent data reduction described in \cite{Cotera&Hankins2024}. We summarize the survey specifications in this section. 

The SOFIA/FORCAST Galactic Center Legacy Survey is a targeted survey designed to image MIR-bright regions of the inner 200 pc of the Milky Way Galaxy using FORCAST filters F252 (centred at 25.2 \micron{}) and F371 (centred at 37.1 \micron{}). The full width at half maximum (FWHM) resolution of the 25 \micron{} observations is $\sim$2.3\arcsec{} and the measured point source sensitivity is 0.75 mJy. At 37 \micron{}, the FWHM is $\sim$3.4\arcsec{} and the measured point source sensitivity is 1.5 Jy. {Overall, the estimated photometric errors for both filters are $\sim10$\%. }

{Survey data were collected during the annual SOFIA southern hemisphere deployment to Christchurch, New Zealand. In cycle 7, 35 fields were observed over eight flights between July 1 and 11, 2019. In cycle 9, 10 more fields (three of which were a part of the Legacy survey but not observed due to scheduling constraints) were observed between July 1, 2021 and May 26, 2022. The integration time across fields varied significantly due to flight conditions. However, the average integration time achieved is about 432 seconds in the 25 \micron{} filter and 416 seconds in the 37 \micron{} filter. The final nominal 5$\sigma$ point source depths are 250 mJy at 25 \micron{} (equivalent to a 3$\sigma$ extended source depth of 1200 ${\rm MJy\,sr^{-1}}$) and 550 mJy at 37 \micron{}. } 

In our initial analysis performed for this work, we found a small background offset between the data collected in Cycles 7 and 9. We therefore applied additional processing from the works of \cite{Hankins2020} and \cite{Cotera&Hankins2024}, with details listed below.

{Noting that the background matching issue is common in FORCAST images when the source emission is more extended than the native instrumental field of view, we implement additional corrections by applying DC offsets to individual frames of the fully calibrated and co-added level 3 data products prior to mosaicking} to bring the background to a comparable level in regions of overlap between frames. {The final DC offset added or subtracted from any given frame is less than the 1$\sigma$ fluctuations of the measured background noise. Such an offset results in backgrounds that are slightly positive, but within 2$\sigma$ of zero \citep[also described in][]{Cotera&Hankins2024}. }

Observations of the Sgr C region also suffer from a { well-known imaging artifact of the FORCAST instrument known as the ``bowl'' effect \citep[e.g., see][]{Lau2013}, in which regions adjacent to  extended, bright sources within a frame can be over-subtracted.} This likely contributes to the above-mentioned background matching issue {for fields near Sgr C. Several regions within} the survey exhibit this effect and are discussed in \citet{Hankins2020}. 

With the need to adjust background levels in individual frames, we rebuild the survey mosaics {from the existing level 3 data products}. The steps for constructing the mosaics presented in this work largely follow from the process detailed in \cite{Cotera&Hankins2024}. {Level 3 data products were retrieved from the NASA/IPAC Infrared Science Archive\footnote{\url{https://irsa.ipac.caltech.edu/frontpage/}} (IRSA; Plan IDs listed previously).} Data processing for these data products uses the \texttt{Redux} software package following the procedure is described in \citet{Clarke2015}. {After supplying the aforementioned DC offset to improve agreement in background levels across adjacent fields, we utilized the \texttt{Reproject} package \citep{Reproject} to create updated mosaics of the Sgr C region and adjacent fields. We compute an optimal world coordinate system for the final mosaic based on the data coverage, and produce the final mosaic using the reproject\_exact() function to conserve flux using a spherical polygon intersection method\footnote{\url{https://reproject.readthedocs.io/en/stable/api/reproject.reproject_exact.html\#reproject.reproject_exact}}. Additional data quality checks were performed prior to and after mosaicking, similar to the process described in \cite{Hankins2020} and \citet{Cotera&Hankins2024}.}

\section{The SOFIA/FORCAST View of Sgr C}
\label{sec:analysis}
In this section, we present a series of analyses of the Sgr C region using the SOFIA/FORCAST mosaic introduced in Section~\ref{sec:obs}. We lead with an overview of the region, then present maps of dust temperature and optical depth, followed by accounts of the detailed properties of FORCAST-detected individual sources. 

\subsection{Overview of the Region}
\begin{figure*}[t!]
    \centering
    \includegraphics[width=\textwidth]{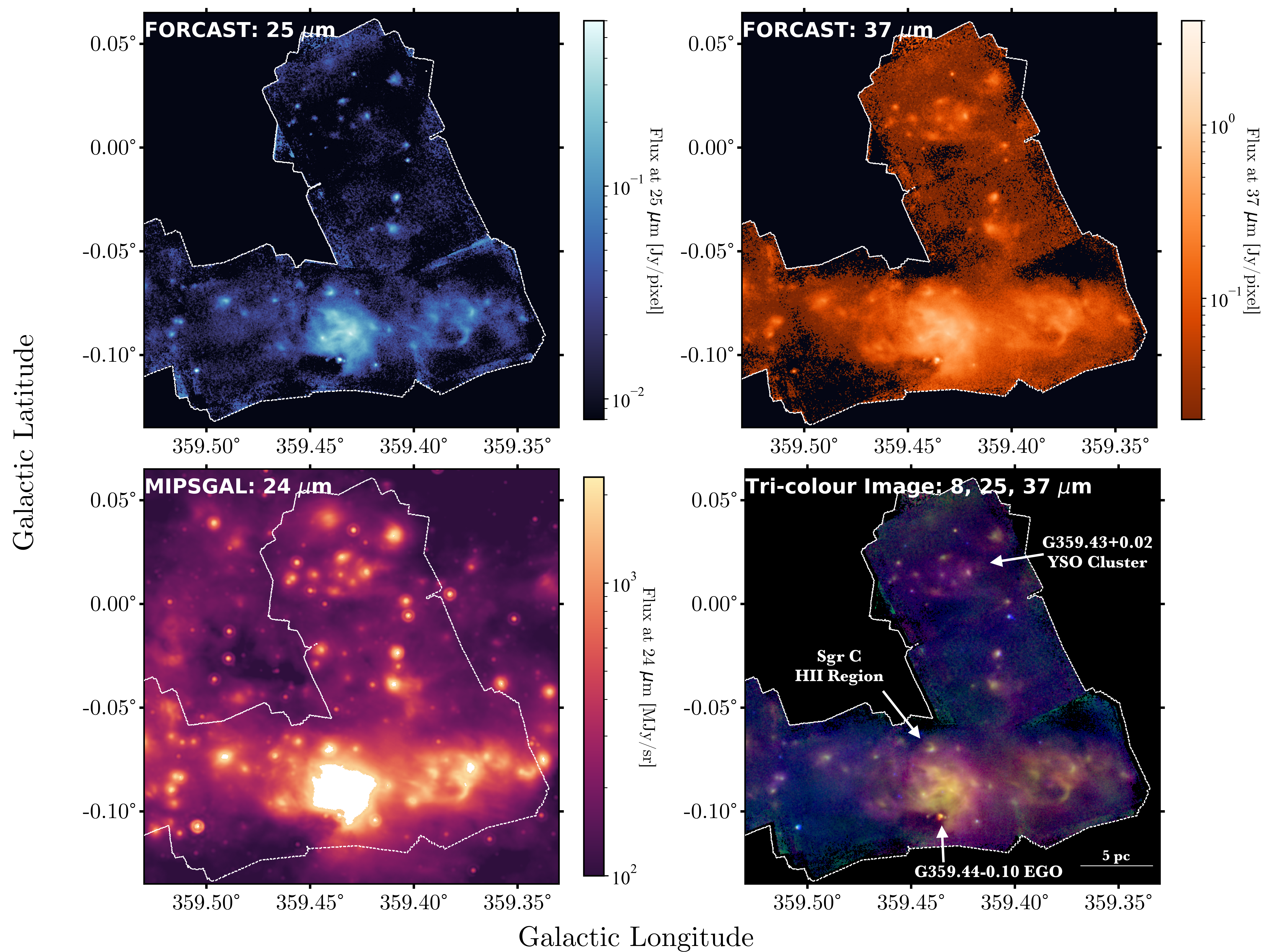}
    \caption{The SOFIA/FORCAST Galactic Center Survey coverage in the Sgr C region at 25 and 37 \micron{} \citep[\textit{top left and top right, respectively};][]{Hankins2020, Cotera&Hankins2024}. The lower left panel displays the {\it Spitzer}/MIPS image at 24 \micron{} \citep[MIPSGAL;][]{Carey2009}. The lower right panel shows a tri-colour rendition using the {\it Spitzer}/IRAC image at 8 \micron{} as blue \citep{Stolovy2006}, and FORCAST 25 and 37 \micron{} images as green and red, respectively. The white outline in each panel denotes the image boundary of the FORCAST survey. The locations of major MIR sources within the region are labelled in the lower right panel. A scalebar of 5 pc length is placed in the lower right panel.}
    \label{fig:comb_plots}
\end{figure*}

Figure~\ref{fig:comb_plots} shows the SOFIA/FORCAST survey fields in the Sgr C region at 25 and 37 \micron{} in the top panels \citep{Hankins2020, Cotera&Hankins2024}, along with the 24 \micron{} {\it Spitzer}/MIPS image from the MIPSGAL survey at the lower left \citep{Carey2009} and a tri-colour depiction combining the {\it Spitzer}/Infrared Array Camera (IRAC) 8 \micron{} image with the FORCAST images at the lower right \citep{Stolovy2006}. 
There are many MIR sources along the Galactic plane, including Source C (labelled in the lower right panel of Figure~\ref{fig:comb_plots}) and the G359.43+0.02 YSO cluster. Globally, the structures appearing at 37 \micron{} are typically more visually diffuse than their 25 \micron{} counterparts, due in part to the lower spatial resolution at 37 \micron{}, but dust heating can also contribute to some extent to the sources being more diffuse at longer wavelengths; dust lying closer to radiation sources (e.g., stars and \hii{} regions) can be heated to a higher temperature, therefore leading to a more compact morphology at shorter wavelengths where higher temperatures dominate. 

{Here, we note the difference in the sensitivities of the MIPS and FORCAST instruments mentioned in the introduction. With its higher sensitivity and lower saturation limit, MIPS provides a superior sampling of the diffuse background of warm dust, but the \hii{} region and its immediate surroundings are saturated. The SOFIA/FORCAST data nicely complement the MIPSGAL data by providing an unsaturated image of the brightest regions, but without such a highly significant measure of the weak, diffuse background. The combination of these two images, therefore, is able to provide us with a complete picture of warm dust in the Sgr C region. }

\begin{figure*}[t!]
    \centering
    \includegraphics[width=\textwidth]{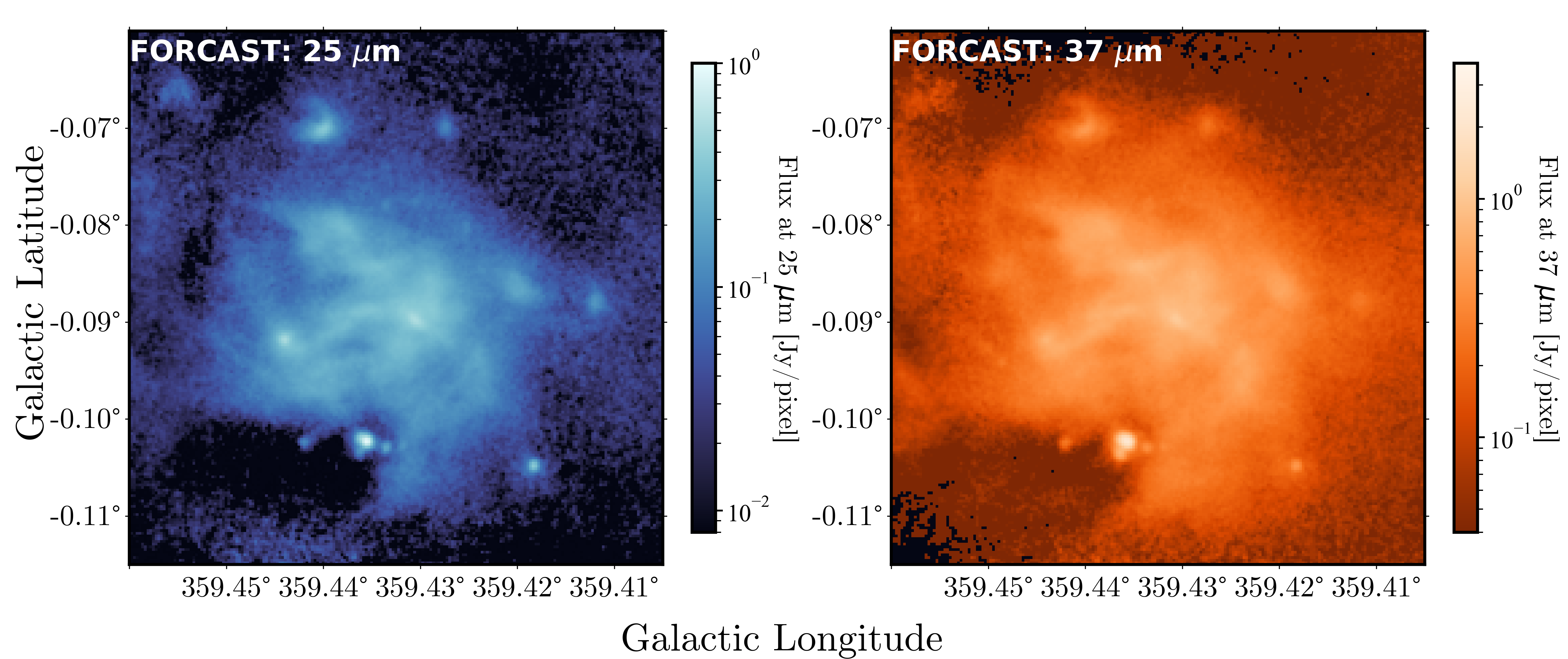}
    \caption{SOFIA/FORCAST images of the Sgr C \hii{} region at 25 and 37 \micron{}, respectively. Both images are displayed with a logarithmic stretch.}
    \label{fig:comb_plots_HII}
\end{figure*}

The most prominent MIR structure in Figure~\ref{fig:comb_plots} is the Sgr C \hii{} region. This \hii{} region has a signature quasi-spherical morphology roughly ten parsecs in diameter, as observed in both MIR and radio continuum \citep{Liszt1995}. Figure~\ref{fig:comb_plots_HII} displays the details of the Sgr C \hii{} region in both of the FORCAST bands.
We focus our discussion on the MIR observation in this section and note that a comprehensive ensemble of multi-wavelength analysis of the Sgr C \hii{} region is presented later in Section~\ref{sec:HII}. In the central portion of the \hii{} region, we observe several overlapping dust emission ridges. \citet{Bally2024} reported the same structures in their JWST-NIRCam observation and argued that they likely trace ionization fronts. These features are further discussed in the context of their dust temperatures and optical depths in Sections~\ref{sec:colourtemp} and \ref{sec:opticaldepth}, respectively.

A number of compact sources lie outside the perimeter of the \hii{} region. All observed features appear more diffuse in the 37 \micron{} image, similar to what is observed globally. We identify the G359.43-0.10 EGO as a very bright and compact MIR source, signifying the substantial heating by the embedded protostellar objects \citep{Kendrew2013, Lu2019, Crowe2024}. As we address later in Section~\ref{sec:discussion}, the absence of MIR emission towards the southeast of the EGO is possibly attributable to absorption by a dense, cold, foreground molecular cloud with a column density of $N(\mathrm{H_2}) \sim 10^{23} \, \mathrm{cm^{-2}}$ that can only be sampled with FIR observations \citep{Zhao2025}. 

Another intriguing source in the region is the G359.43+0.02 cluster, from where \citet{Yusef-Zadeh2009} reported 18 confirmed YSOs. This cluster has a counterpart in the FIR, first identified and named by \citet{Odenwald1984} as FIR-4. It coincides with a linear ridge of radio emission located to the north of the \hii{} region \citep{Roy2003, Lang2010}, which might be powered by emission from the cluster. The {\it Spitzer}/MIPS 24 \micron{} image of this region shows a wealth of compact sources along with diffuse, extended dust features, similar to the morphology observed by FORCAST at 37 \micron{}. However, the FORCAST 25 \micron{} image predominantly features the brightest of the cluster sources, as much of the diffuse emission observed by MIPS is below the sensitivity threshold of the FORCAST survey at this wavelength. We return to a discussion of this cluster in Section~\ref{sec:YSO}.

\subsection{MIR Analysis}
\label{sec:T+tau}
The images obtained with FORCAST at 25 and 37 \micron{} are dominated by dust heated by the interstellar radiation field \citep{Hankins2020}. The combination of observations at these two wavelengths enables us to derive dust temperature and optical depth maps of the Sgr C region. 

\subsubsection{Extinction \& Signal-to-noise Correction}
\label{sec:extinction}
In order to draw meaningful physical inferences from the luminosities of objects in the GC, we must first correct for the substantial foreground extinction \citep{Weingartner2001, Nishiyama2008, Fritz2011}. {To account for spatially varying extinction in different regions}, we utilize the dust extinction map reported by \citet{Simpson2018} from {\it Spitzer}/IRS observations in the CMZ, presented as an optical depth map at 9.8 \micron{}, which we denote as $\tau_{\rm sil}$ to highlight that this wavelength is used to trace a silicate feature\footnote{To avoid confusion, we note that there are a few wavelength conventions in the literature ranging from 9.6 to 9.8 \micron{} to trace the same underlying silicate feature. We therefore adopt a general notation $\tau_{\rm sil}$ to denote the optical depth of this feature.}. Hereafter, we denote the optical depth at wavelength $\lambda$ \micron{} to be $\tau_\lambda$. {Since the {\it Spitzer}/IRS observations are not spatially uniform across the entire CMZ, we adopt an interpolated version of the $\tau_{\rm sil}$ map made using the nearest neighbours algorithm (provided by J. P. Simpson; private communication) {with a 6\arcsec{} native pixel size, then convolved using a 2-pixel $\sigma$ to smooth the sharp boundaries between different observation patches}. Although this map does not explicitly embed extinction towards individual sources, the high sample density within the Sgr C region provides a resolution sufficient to distinguish the extinction variations between individual sources and features.} 

To account for the extinction as a function of wavelength, we adopt the {MIR-specific} extinction curve from \citet{Chiar2006} to {extrapolate} the 9.8 \micron{} optical depth at a given position to the continuum bands measured by the FORCAST filters centered at 25.2 and 37.1 \micron{}, giving $\tau_{25.2}/\tau_{\rm sil}=0.41$ and $\tau_{37.1}/\tau_{\rm sil}=0.28$. {In the Sgr C region, there are a few dense, cold molecular clouds \citep{Kendrew2013, Zhao2025} that are not properly sampled with the FORCAST camera. As this paper focuses on the MIR regime, we determine our extinction corrections using MIR observations only. }

Due to the spatial resolution difference at the two FORCAST wavelengths, we convolve both {mosaicked} FORCAST images to a lower 5\arcsec{} resolution using a 2D Gaussian kernel before computing the dust temperature {while keeping the original FORCAST plate scale}. The FWHM of the said Gaussian kernel is computed as 
\begin{equation}
    {\rm FWHM}_{G, \lambda} = \sqrt{W_{\rm Target}^2 - W_{\lambda}^2},
\end{equation}
where $W_{\rm Target}=5$\arcsec{} is the resolution of the final convolved maps and $W_{\lambda}$ is the resolution of the FORCAST images at wavelength $\lambda$. 

Since the survey data were obtained over the span of a few years, the frame-to-frame noise properties of the FORCAST images vary non-trivially due to differences in integration time and observing conditions. As such, we cannot assume a uniform signal-to-noise ratio (SNR) throughout the mosaicked images. Instead, we assigned appropriate SNR cutoffs to individual frames prior to mosaicking. This method provided the optimal trade-off between excluding any pixel below the locally measured $3\sigma$ level, and preserving emission in frames that might have been cut off if a uniform SNR cutoff across all frames were adopted. Although some information contained in diffuse emission is lost, a higher SNR cutoff enables an accurate calculation of the physical properties of luminous MIR regions. We adopt the mosaic processed in this fashion throughout the rest of Section~\ref{sec:analysis}.

{Throughout this paper, we do not explicitly consider stochastic heating of very small dust grains (VSGs) and its effect on the MIR wavelengths. Previous works on various GC \hii{} regions found that the emission at MIR wavelengths is consistent with grains in thermal equilibrium \citep{Hankins2017, Hankins2019}. In the SED modelling we perform in Section~\ref{sec:totlum}, we show that the emission from PAHs and VSGs (that are stochastically heated grains) are not the dominant species at 25 and 37 \micron{}. However, the effect of stochastic heating is important in sources that are distant from their heating source such that the photon hit rate is low \citep[e.g., observations of extragalactic objects; ][]{Relano2016}.}

\subsubsection{Dust temperature map}
\label{sec:colourtemp}
\begin{figure*}
    \centering
    \includegraphics[width=\textwidth]{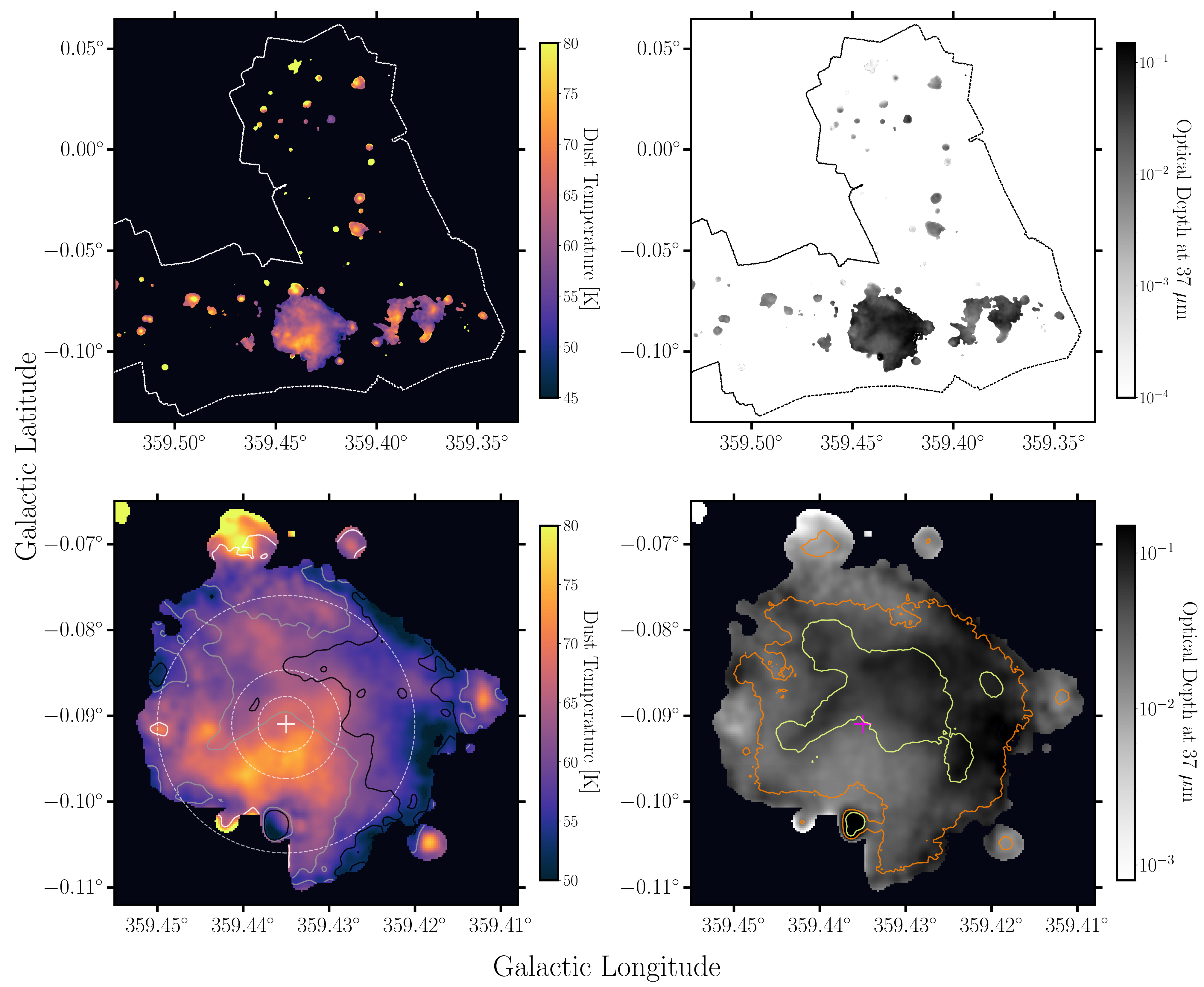}
    \caption{{\it Top panels:} The map-level, per-pixel dust temperature ({\it left}) and optical depth at 37 \micron{} ({\it right}) in the Sgr C region made from the SOFIA/FORCAST data. {\it Bottom left:} The dust temperature map zoomed into the Sgr C \hii{} region overlaid with 37 \micron{} optical depth contours at (0.01, 0.04, 0.07) with higher optical depths depicted in darker shades. The white cross indicates the location of a spectroscopically-confirmed WCL star referred to in this work as the central WCL star \citep{Geballe2019, Clark2021}. The dashed white circles are the heating profiles of the central WCL star drawn at {[60, 80, 100] K, with radii of [54.0, 22.8, 11.7]\arcsec or [2.15, 0.91, 0.47] pc}. The detailed calculation is presented in Section~\ref{sec:tempprofile}. {\it Bottom right:} The 37 \micron{} optical depth zoomed into the Sgr C \hii{} region overlaid with 37 \micron{} flux density contours at (0.25, 0.375, 0.5) Jy/arcsec$^2$ with higher temperature depicted with brighter shades. The magenta cross again labels the central WCL star. }
    \label{fig:combined_T_tau}
\end{figure*}

In our dust temperature calculation, we assume the dust emissivity follows a {modified blackbody model}, where $F_\nu \propto \nu^\beta B_\lambda$, such that the flux $F_\nu$ measured at frequency $\nu$ follows a blackbody $B_\nu$ weighted by a frequency-dependent power-law emissivity with index $\beta$. Here, we adopt $\beta=2$ {following \cite{Draine2011} for a silicate-dominated warm dust population\footnote{which is further confirmed by the SED model shown in Section~\ref{sec:totlum}}, similar to the FORCAST analysis by \citet{Hankins2017}}. The dust temperature at each pixel, $T_d$, is therefore obtained by solving 
\begin{equation}
    \frac{F_{25}}{F_{37}} = \left(\frac{\lambda_{37}}{\lambda_{25}}\right)^7 \frac{{\rm exp}\left(\frac{hc}{\lambda_{37}k_{\rm B}T_{\rm d}}\right)-1}{{\rm exp}\left(\frac{hc}{\lambda_{25}k_{\rm B}T_{\rm d}}\right)-1}
    \label{eq:fluxratio}
\end{equation}
in wavelength space, where $F_\lambda$ is the extinction-corrected flux at one of the FORCAST-observed wavelengths $\lambda$ denoted by the subscript, $h$ is the Planck constant, $k_B$ is the Boltzmann constant, and $c$ is the speed of light. The top left panel of Figure~\ref{fig:combined_T_tau} shows the computed dust temperature map of the full Sgr C region, while the lower left panel shows a zoomed-in view of the Sgr C \hii{} region. 
The MIR-bright regions commonly have temperatures between 60--80 K. The Sgr C \hii{} region exhibits a lower temperature of $\sim 55-70$ K, which is lower than several sources to the east of the \hii{} region and in the G359.43+0.02 YSO cluster, which have typical temperatures higher than 70 K. All of these compact sources are classified in the FORCAST source catalogue \citep{Cotera&Hankins2024} and are found to have 8 \micron{} counterparts in {\it Spitzer}/IRAC images, therefore they are likely stellar sources. To the west of the \hii{} region, Source C exhibits a temperature similar to that of the \hii{} region, with some high temperature knots that range up to $\sim 70$ K.

The zoomed-in dust temperature map of the Sgr C \hii{} region shows that the periphery of the \hii{} region towards the west and the G359.44-0.10 EGO have the lowest dust temperature $\sim 50$ K. The highest temperature region, with $\sim 70$ K, occurs in its southern portions, which is located to the south of the previously mentioned dust ridges (see the continuous grey optical depth contours in the lower left panel of Figure~\ref{fig:combined_T_tau}; detailed in Section~\ref{sec:opticaldepth}) and a confirmed Wolf-Rayet star immediately below the ridges (white cross in the same Figure; detailed in Section~\ref{sec:tempprofile}). The dust ridges do not stand out in the dust temperature map. However, when compared with the optical depth contours, the high-temperature region appears to be contained below the ridges. 

{We note a caveat that the short wavelength spacing of the FORCAST filters (between 25 and 37 \micron{}) can lead to poor dust temperature constraints in high temperature regions. Given our modified blackbody assumption, a temperature higher than $T_{\rm d}\sim82$ K shifts the peak wavelength shorter than $\sim25$ \micron{}, so that both wavelengths are then placed on the longward side of the SED peak. In that region, the flux ratio is  insensitive to temperature variations. In our analysis, a few compact sources (e.g., the compact source at $l=359.44$\degree{}, $b=0.07$\degree{}) exhibit temperatures close to this limit, so their temperature estimation should be taken with caution. }

\subsubsection{Optical Depth map}
\label{sec:opticaldepth}
An optical depth at these wavelengths can be derived in concert with the computed dust temperature. Following the assumption that the dust is optically thin, which is widely adopted in studying the CMZ \citep[e.g.,][]{Hankins2019}, the optical depth at a given wavelength $\lambda$ at each pixel is approximated as 
\begin{equation}
    \tau_\lambda = \frac{F_\lambda}{\Omega_p B_\lambda(T_d)},
    \label{eq:tau}
\end{equation}
where $\Omega_p$ is the solid angle subtended by a single FORCAST pixel. We show the global optical depth map calculated at 37 \micron{} in the top right panel of Figure~\ref{fig:combined_T_tau}. We find that the highest optical depth globally has values of $\sim$0.1, supporting the thin dust approximation. Regions with high MIR luminosity (e.g., the Sgr C \hii{} region) are found to have a typical optical depth of $\sim 0.1-0.01$, whereas compact sources in the G359.43+0.02 YSO cluster and to the east of the \hii{} region are found to have relatively low optical depth of $< 0.01$, in accordance with the conclusion that they are stellar or young stellar objects situated in a low-density ISM. The G359.44-0.10 EGO has the highest optical depth of $> 0.1$, again confirming the high cloud density found in previous studies \citep{Kendrew2013, Crowe2024}, which favours massive star formation. 

{For optical-depth-related analysis throughout this work, we note that the MIR-derived optical depth value of these high-density molecular clouds is likely not well represented. As the emission from these clouds is predominantly in the FIR (e.g., various {\it Herschel} observations; \citealt{Molinari2010, Battersby2025}), their dust opacity is poorly sampled by MIR observations with FORCAST, which could be dominated by thin layers of warm dust emission.} 

The bottom right panel of Figure~\ref{fig:combined_T_tau} presents a zoomed-in view of the map-level optical depth of the \hii{} region overlaid with the 37 \micron{} flux density contour. The highest optical depth is found in the central portion of the \hii{} region, with values as high as $\gtrsim 0.1$. The emission from these dust ridges produces the emission ridges previously identified in the MIR intensity images (roughly indicated by the extent within the innermost orange contour in the lower right panel of Figure~\ref{fig:combined_T_tau}, c.f., Figure~\ref{fig:comb_plots_HII}). Therefore, we conclude that these structures are enhanced-density dust ridges, as the elevated MIR luminosity is likely due to the high optical depth as a proxy for high column density. The dust temperature and optical depth are further discussed in subsequent sections. {Additionally, as optical depth is a function of dust temperature, we again note that the optical depths of high-temperature compact sources (with $T\sim82$K) should be taken with caution. }

\subsubsection{Temperature Profile Modelling}
\label{sec:tempprofile}
Recently, \citet{Geballe2019} and \cite{Clark2021} reported two spectroscopically-determined late-type, carbon-sequence Wolf-Rayet stars (hereafter denoted WCL stars, based on their spectroscopic characteristics) in the Sgr C complex. Both are confirmed by \cite{Oka2019} to be in the CMZ, with one of them suggestively located in projection at the centre of the Sgr C \hii{} region (cross symbol in the bottom panels of Figure~\ref{fig:combined_T_tau}). It is therefore of great interest to study their effect on the surrounding ISM. Between these two WCL stars, the one located outside the \hii{} region has no observable association with other physical features across all wavelengths. Therefore, we dedicate the subsequent discussion almost exclusively to the WCL star at the projected centre of the \hii{} region (2MASS label J17443734-2927557), which we hereafter denote as the central WCL star unless otherwise noted.

Here, we estimate the heating effect of the central WCL star on the surrounding ISM and evaluate whether this star can be a primary source of heating. We adopt the model from \citet{Hankins2019} for calculating the heating profile of the ISM. This method provides an estimate of the maximum temperature of the dust at a given distance from a compact heating source, which in our case is the central WCL star, assumed to be isolated. The modelled temperature profile can then be compared with the dust temperature map from Section~\ref{sec:colourtemp} to infer the astrophysical properties of the ISM. In our calculation, we assume the dust grains are in thermal equilibrium with the incident and unextinguished radiation field emanating from the stellar source with luminosity $L_\star$. The dust temperature as a function of distance $l$ from the source is computed as
\begin{equation}
    T_d(l) = \left(\frac{L_\star}{16\pi\sigma_{\rm SB} l^2}\frac{Q_{\rm UV}}{Q_d'}\right)^{1/(\beta+4)},
    \label{eq:T_d}
\end{equation}
where $\sigma_{\rm SB}$ is the Stefan-Boltzmann constant, and $\beta$ is the dust emissivity exponent, and we adopt the same value, $\beta=2$, as was used for our dust temperature calculation. The $Q_{\rm UV}$ factor is the grain absorption cross section averaged over the spectrum of the incident stellar radiation field, expressed as an integral over the frequency-dependent cross-section of the grain $Q_{\rm abs}(\nu)$, where
\begin{equation}
    Q_{\rm UV} = \frac{\int d\nu B_\nu(T)Q_{\rm abs}(\nu)}{\int d\nu B_\nu(T)}.
    \label{eq:Q_UV}
\end{equation}
Furthermore, the last unintroduced parameter $Q_d'$ in Equation~\ref{eq:T_d} is the temperature-independent dust emission efficiency. We adopt $Q_d'=1.3\times10^{-6}(a/0.1~{\rm \mu m})$ from \citet{Draine2011} as the expression for silicate grains, and a characteristic grain radius of $a=0.01$ \micron{} in this calculation. Although lower than the typical ISM dust grain size of $\sim0.1$ \micron{}, this choice is consistent with the previously determined dust grain properties of various \hii{} regions in the CMZ \citep{Lau2016, Hankins2017, Hankins2019}. As part of our study, we have attempted the calculation using the typical ISM grain size, $a=0.1$ \micron{}, but the temperature profiles are much smaller in that case and provide a poorer fit to the observed dust temperature map, possibly suggesting the existence of smaller grains in the Sgr C \hii{} region. 

{We further correct for the projection effect due to the unknown line-of-sight offset of the heating source. For photometric surveys, we only observe the plane-of-sky source-target offset $R_{\rm POS}$ but not the line-of-sight offset $R_{\rm offset}$, while the true 3D distance $l$ is related to both via $l = \sqrt{R_{\rm POS}^2+R_{\rm offset}^2}$. To convert $R_{\rm POS}$ to a characteristic 3D distance $l_{\rm c}$, we adopt an isotropic assumption that $R_{\rm POS}\approx R_{\rm offset}$, therefore obtaining $l_{\rm c}\approx\sqrt{2}R_{\rm POS}$. In the subsequent discussion, we adopt only this characteristic distance. }

The circular dashed contours in the lower left panel of Figure~\ref{fig:combined_T_tau} display the calculated radial dust heating profile at (60, 80, 100) K around the central WCL star progressing from larger to smaller radii, assuming its luminosity to be $10^{39}$ erg/s. We observe that the modelled profile shows higher temperatures than those observed in regions close to the source (within the 80 and 100 K contours), since the observed temperature there has a maximum of $\sim75$ K, possibly in part due to the smoothing effect of applied convolution detailed in Section~\ref{sec:extinction}. However, further from the source in the annulus between the 60 and 80 K contours, the observed temperature reasonably fits the calculated temperature profile. We also note that the observed temperature in the southern portion of the $60$ K circle is much higher than that in the northern portion. the agreement between the heating profile and the dust temperature map suggests that the central WCL star could be the dominant contributor, or that any additional luminous stars that might be present are located close to the WCL star, subject to the uncertainty of grain size in this region. 

The optical depth map in the lower right panel of Figure~\ref{fig:combined_T_tau} shows the location of the X-shaped dust ridges to the immediate north of the central WCL star (magenta cross), and a relatively uniform region south of it. Therefore, the elevated density in the dust ridges to the north may cause a higher rate of ISM cooling relative to the heating by the star, assuming the dust is collisionally coupled to the gas. It is also possible that the winds of the WCL star could be slowed down by the high-density ISM, thus compressing the local gas and inducing the formation of higher-density ridges. 

\subsubsection{Point Sources}
\begin{figure*}[t!]
    \centering
    \includegraphics[width=\textwidth]{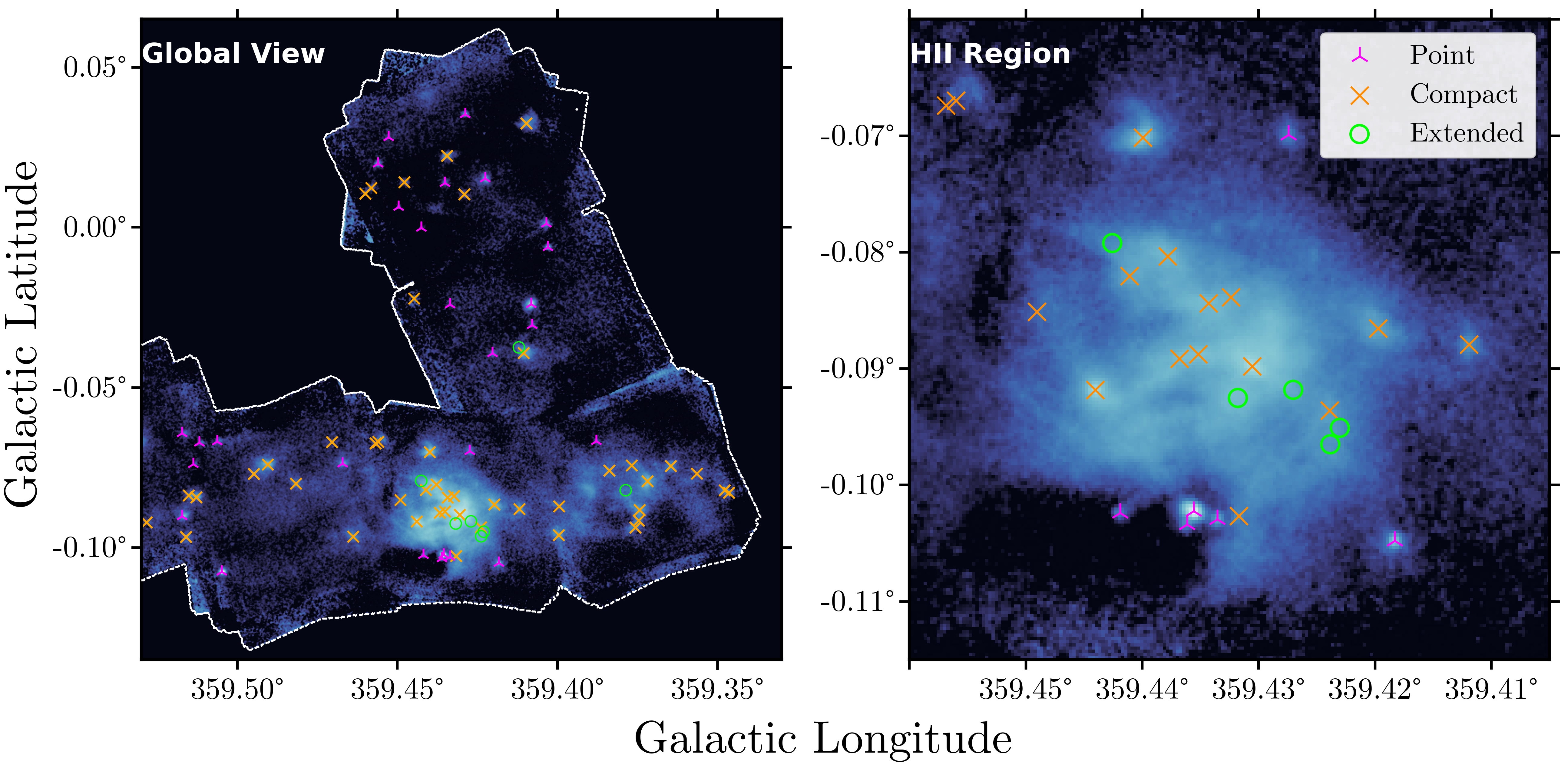}
    \caption{The categorized FORCAST-detected individual sources overlaid on the 25 \micron{} image \citep{Hankins2020, Cotera&Hankins2024}. The left panel shows the full observed Sgr C complex, and the right panel zooms into the \hii{} region. The magenta markers, orange crosses, and green circles label point, compact, and extended sources, respectively. }
    \label{fig:ptsourceimage}
\end{figure*}

We here incorporate the FORCAST individual source catalogue {published by \citet{Cotera&Hankins2024} into our analysis. It contains 950 sources found CMZ-wide in the SOFIA/FORCAST survey map. The point sources in this catalogue were extracted using the DAOFIND algorithm \citep{Stetson1987} and the image segmentation technique included in the \texttt{Photutils} package \citep{Bradley2020} as the DAOStarFinder routine. The photometry of the extracted sources was calculated using multiple methods, and sources with $\text{SNR}<3$ were removed.} Before drawing any formal conclusion, it is necessary to note that this catalogue includes a wide variety of MIR sources, ranging from unresolved point sources, resolved but compact sources, and extended sources. Many of the sources also have complex morphology and background structure, so a simple classification of the sources based on their FWHM measurements is likely heavily contaminated by the complex background emission (e.g., examples shown in Figures 4 and 5 in \citealt{Cotera&Hankins2024}). {For this reason, the FORCAST source catalogue offers two avenues of classifying the sources: i) a traditional 2D Gaussian and 2D Moffat fit, and ii) an empirically-motivated method using the ratios of the fluxes measured in differently sized apertures around the sources, defined as the pixel flux ratio. The former is more accurate for classifying point-like sources, and the latter is found to be more accurate for sources with complex background emission. We have tested both avenues, and choose to adopt the latter due to the substantial background emission in Sgr C, especially in the \hii{} region. 
We classify the sources found in the Sgr C region using two measures of source extent: the 8/4 and 12/8 pixel flux ratios. Here, the $x$/$y$ pixel flux ratio of a given source is the quotient of flux measured with circular aperture $x$ pixels in radius divided by flux measured with an aperture $y$ pixels in radius.} We classify our sources into three categories: 
\begin{itemize}
    \item Point sources, defined by 8/4 pixel flux ratio $<2.0$ and 12/8 pixel flux ratio $<1.3$
    \item Compact sources, defined by 8/4 pixel flux ratio $<3.0$ and 12/8 pixel flux ratio $<1.8$, excluding point sources
    \item Extended sources, defined by 8/4 pixel flux ratio $\geq3.0$ and 12/8 pixel flux ratio $\geq1.8$
\end{itemize}
The spatial distribution of the classified sources is displayed in Figure~\ref{fig:ptsourceimage}.  
{Given the conjunction conditions, there are a total of 26 sources that remain unclassified in the region around Sgr C (field-of-view of the left panel of Figure~\ref{fig:ptsourceimage}), with 10 of those located within the Sgr C \hii{} region (field-of-view of the right panel of Figure~\ref{fig:ptsourceimage}). Our visual inspection of the unclassified sources revealed that most of them have irregular morphologies, except for three fainter, more compact sources near the bright G359.44-0.10 EGO core\footnote{These unclassified sources are indexed as SFGC359.442-0.1024, SFGC359.434-0.1029, SFGC359.436-0.1034 in the \citet{Cotera&Hankins2024} catalogue using Galactic coordinates.}. Since these three sources are visually more compact than the EGO core (which is classified as a point source), we manually classify these three sources as point sources as well. We do not consider any other unclassified sources in the subsequent discussions.} Globally, we observe that although there are a few point sources on the periphery of the Sgr C \hii{} region (e.g., the G359.44-0.10 EGO), it is largely devoid of point sources. Most of the point sources are within the G359.43+0.02 cluster and further east along the Galactic plane, where no extended sources by the above definition are present. Within the \hii{} region, we found many ``compact" and ``extended"  sources that are spatially distributed along the dust ridges that we observe within the \hii{} region. These sources are the bright MIR peaks that are present along the dust ridges. 

As mentioned earlier, \citet{Cotera&Hankins2024} provide the modelled fluxes using multiple photometric fitting methods, for each source. With fluxes at two wavelengths, one can in principle conduct a source-level temperature and optical depth analysis similar to what we report in Section~\ref{sec:analysis}. However, many point or compact sources are found with 8 \micron{} {\it Spitzer}/IRAC counterparts and therefore are likely the dust re-emission from stellar sources. The validity of our dust temperature calculation becomes unclear in small-scale environments, especially after convolution. Therefore, {we do not attempt to calculate the dust temperature and optical depth of individual sources using their modelled fluxes provided by the FORCAST source catalogue, but rather adopt these quantities from the sources' locations from the map-level dust temperature and optical depth calculations presented in Figure~\ref{fig:combined_T_tau} as the sources' temperature and optical depth. However, since the catalogued sources are extracted from mosaics \textit{without} the additional processing outlined in Section~\ref{sec:extinction} (i.e., the original mosaic shown in Figure~\ref{fig:comb_plots}), some sources in low-SNR regions do not meet the SNR criteria and as such do not have a corresponding temperature or optical depth at those locations from Figure~\ref{fig:combined_T_tau}. These sources exist outside of the \hii{} region and the YSO cluster (except for the source FIR-4, which is discussed below in Section~\ref{sec:YSO}) and are therefore outside of the focus of this work and do not influence our conclusions. 
} 

\section{Multi-wavelength Views of Sgr C}
\label{sec:discussion}
The Sgr C complex has been observed as part of many other Galactic surveys. By combining the FORCAST observations with these data, we are able to perform a more comprehensive analysis of this region. 

\subsection{Assembly of Multi-wavelength Observations}
\begin{figure*}[h!]
    \centering
    \includegraphics[width=\textwidth]{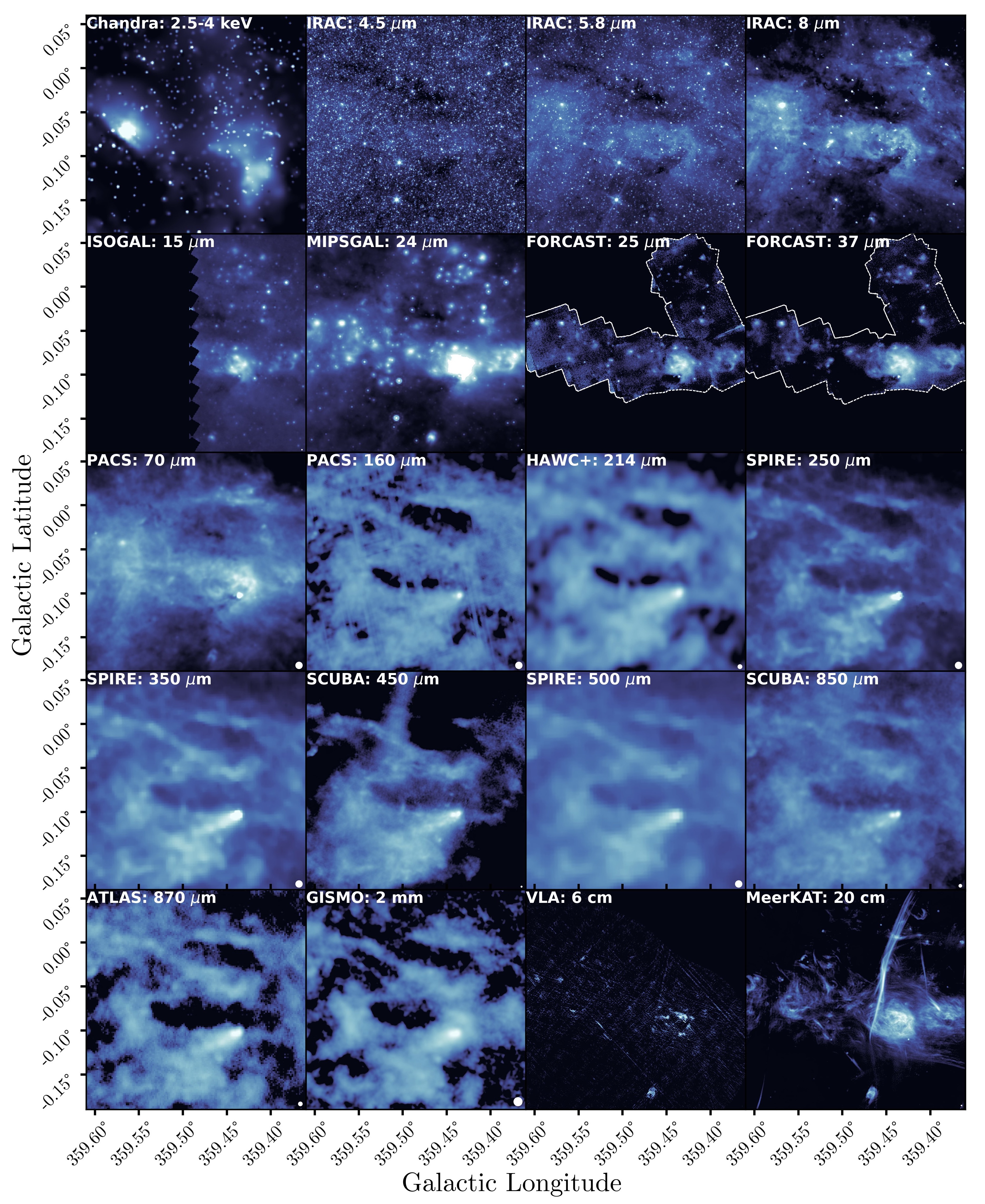}
    \caption{An ensemble of Galactic surveys covering the Sgr C region. Details of the included surveys are presented in Table~\ref{tab:surveys}. The corresponding survey name and wavelength of each image are displayed in the top left corners of individual images. The beam size for each survey is displayed in the lower right corner of each panel. All images are displayed with a logarithmic stretch. }
    \label{fig:sgr_allfields}
\end{figure*}

Figures~\ref{fig:sgr_allfields} and \ref{fig:sgr_allfields_HII} present images of the intensity maps across twenty wavelengths collected by past surveys covering Sgr C, with Figure~\ref{fig:sgr_allfields_HII} illustrating the \hii{} region in detail. The figures include X-ray, near-IR (NIR), MIR, FIR, sub-millimetre, and radio continuum images in the order listed in Table~\ref{tab:surveys} with survey details. {All data products are taken from publicly available archives provided by the referenced authors. Furthermore, we note that there are other existing, high-quality data products from the community \citep[e.g.,][]{Battersby2020, Tang2021}, however, we only select a subset here to display a quasi-continuous picture of the Sgr C region across the electromagnetic spectrum.}

\begin{table*}[t!]
\centering
\caption{Information on the surveys covering Sgr C and presented in Figures~\ref{fig:sgr_allfields} and \ref{fig:sgr_allfields_HII}, {ordered} by the surveyed wavelength.}
\begin{tabular}{lllll}
\hline\hline
Survey & Wavelength & Resolution & Telescope & Reference \\ 
\hline
Chandra & $2.5-4 ~{\rm keV}$ & 0.5\arcsec & {\it Chandra} & \citet{Wang2021} \\
IRAC & 4.5 \micron{} & 2\arcsec & {\it Spitzer} & \citet{Stolovy2006, Arendt2008} \\
IRAC & 5.8 \micron{} & 2\arcsec & {\it Spitzer} & \citet{Stolovy2006, Arendt2008} \\
IRAC & 8 \micron{} & 2\arcsec & {\it Spitzer} &\citet{Stolovy2006, Arendt2008} \\
\hline
ISOGAL & 15 \micron{} & 3\arcsec & {\it ISO} &\citet{Omont2003, Schuller2006}  \\
MIPSGAL & 24 \micron{} & 6\arcsec & {\it Spitzer} & \cite{Carey2009} \\
FORCAST & 25 \micron{} & 2.3\arcsec & SOFIA &\citet{Hankins2020} \\
FORCAST & 37 \micron{} & 3.4\arcsec & SOFIA &\citet{Hankins2020} \\
\hline
PACS & 70 \micron{} & 6\arcsec &  {\it Herschel} &\citet{Molinari2010, Molinari2016} \\
PACS & 160 \micron{} & 12\arcsec &  {\it Herschel} &\citet{Molinari2010, Molinari2016} \\
HAWC+ & 214 \micron{} & 19.6\arcsec & SOFIA &\citet{Pare2024} \\
SPIRE & 250 \micron{} & 18\arcsec & {\it Herschel} &\citet{Molinari2010, Molinari2016} \\
\hline
SPIRE & 350 \micron{} & 24\arcsec & {\it Herschel} &\citet{Molinari2010, Molinari2016} \\
SCUBA & 450 \micron{} & 8\arcsec & JCMT &\citet{Pierce-Price2000} \\
SPIRE & 500 \micron{} & 35\arcsec & {\it Herschel} &\citet{Molinari2010, Molinari2016} \\
SCUBA & 850 \micron{} & 15\arcsec & JCMT & \citet{Pierce-Price2000} 
\\
\hline
ATLASGAL & 870 \micron{} & 19.2\arcsec & APEX & \citet{Schuller2009} \\
GISMO & 2 mm & 21\arcsec & IRAM & \citet{Arendt2019} \\
VLA & 6 cm &  7.5\arcsec & VLA &  \citet{Lu2019}\\
MeerKAT & 20 cm & 4\arcsec & MeerKAT &\citet{Heywood2022}  \\
\hline\hline
\end{tabular}
\label{tab:surveys}
\end{table*}

We first examine global observations of the Sgr C region before detailing the \hii{} region in Section~\ref{sec:HII}. In the shorter-wavelength images that span the first two rows and the leftmost image on the third row, which include X-ray to MIR and then the {\it Herschel}/PACS 70 \micron{} image, the Galactic plane appears bright, illuminated by stellar objects and warm dust, with the Sgr C \hii{} region being the most luminous source in this wavelength regime. In NIR and MIR, the {\it Spitzer}/IRAC images capture the Rayleigh-Jeans emission from stellar sources and polycyclic aromatic hydrocarbon (PAH) emission up to 8 \micron{}. Therefore, images taken at 4.5 and 5.8 \micron{} show abundant stellar sources. The 8 \micron{} image displays continuous structures with the \hii{} region and the Galactic plane as the main features. The MIR images across the second row from 15 to 37 \micron{} display thermal emission of warm dust \citep{Carey2009}, with observed structures similar to those in the 8 \micron{} image. The G359.43+0.02 YSO cluster and source C are also consistently visible in the MIR. 

The FIR structures (from the {\it Herschel}/PACS 160 \micron{} image in the third row to the ATLAS 870 \micron{} image in the bottom row of Figure~\ref{fig:sgr_allfields}) have a very different appearance, as the thermal emission from cold dust now dominates and the heated \hii{} region now becomes a silhouette. The molecular clouds adjacent to the Sgr C \hii{} region, apparently belonging to the 100 pc twisted ring \citep{Molinari2011}, are the most prominent structures at these wavelengths. Additionally, the G359.45-0.10 EGO appears exceptionally bright. The cloud containing this EGO extends a cometary-like tail of cold dust that blends into a large molecular cloud in the south of the \hii{} region. 

Longer than the millimetre scale, the 2 mm GISMO image shows both thermal radiation from cold dust and free-free emission from ionized gas \citep{Arendt2019}. As such, the Sgr C \hii{} region and the cold molecular clouds are both visible. In the radio continuum shown in the last two images of the last row, the emission is entirely dominated by free-free radiation. The 6 cm image is made with a logarithmic stretch to show the brightest portions of the radio structure, and does not capture all of the extended flux. At 20 cm, the \hii{} region and the extended NTF appear as the dominant structures. 

\subsection{The Sgr C \hii{} Region}
\label{sec:HII}
\begin{figure*}[t!]
    \centering
    \includegraphics[width=\textwidth]{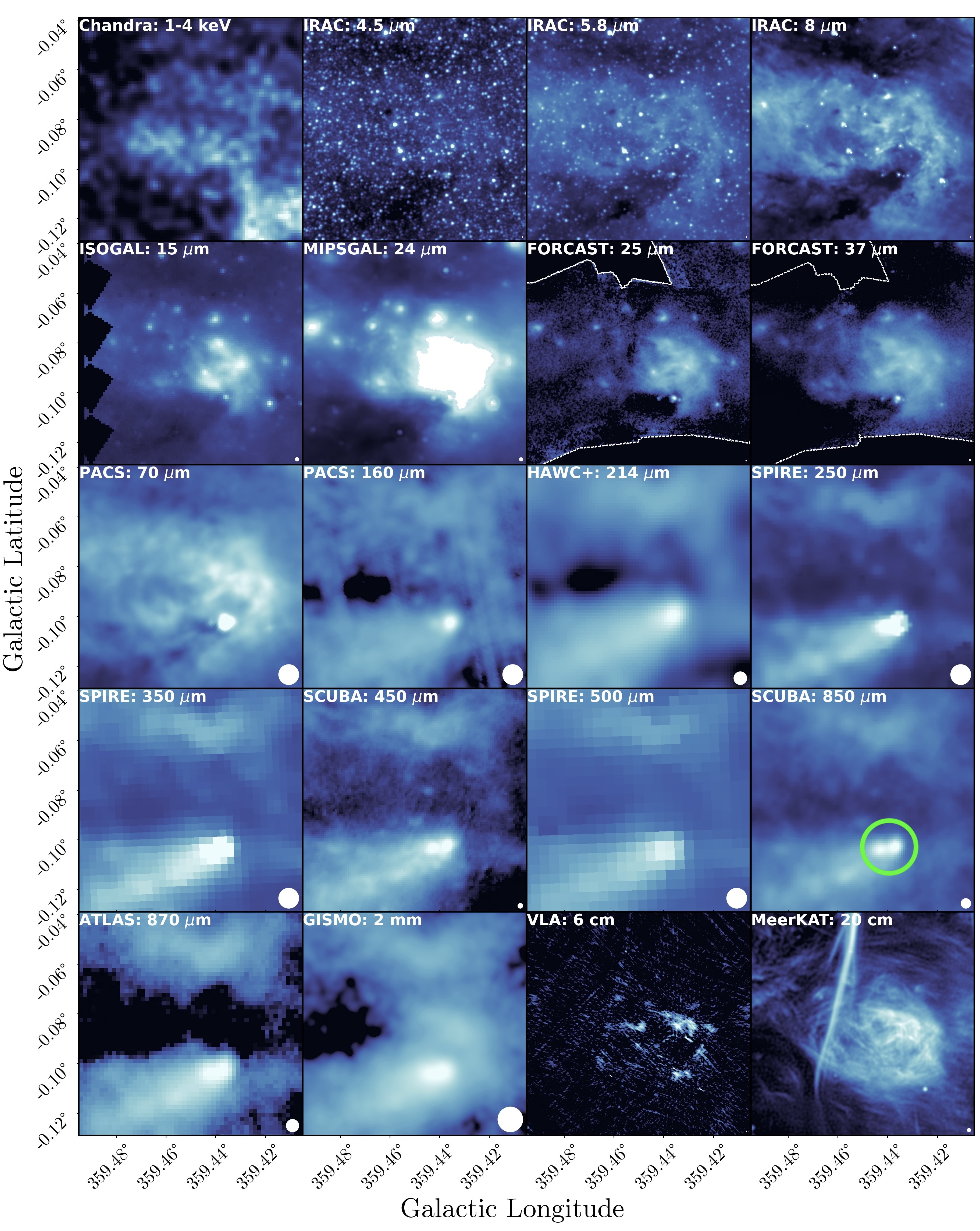}
    \caption{The same ensemble of surveys as Figure~\ref{fig:sgr_allfields}, but zoomed into the Sgr C \hii{} region. {The green circle in the SCUBA image located in the last panel of the fourth row locates the G359.44-0.10 EGO.} }
    \label{fig:sgr_allfields_HII}
\end{figure*}

We turn now to the zoomed-in views of the Sgr C \hii{} region shown in Figure~\ref{fig:sgr_allfields_HII}. 
In the X-ray regime, there is a concentration of 2.5-6 keV X-ray emission observed by \citet{Tsuru2009}, \citet{Wang2021}, and \citet{Zhu2026} that is spatially coincident with the \hii{} region, and an adjacent, even brighter extension of X-ray emission to its southwest, recently discussed by \citet{Zhao2025} and \citet{Zhu2026}. In the NIR images from {\it Spitzer}/IRAC up to 5.8 \micron{}, the emission is dominated by individual stars and only some extended structure is observed. However, the 8 \micron{} image shows prominent PAH emission -- a tracer of ambient UV radiation field -- in an elliptical shape, more extended along the Galactic plane than in MIR and radio images. Further into the MIR (second row of Figure~\ref{fig:sgr_allfields_HII}), the Sgr C \hii{} region progresses into a more circular morphology in the sky-plane. A comparison between the {\it Spitzer}/IRAC image at 8 \micron{} and other MIR images (e.g., {\it Spitzer}/MIPS at 24 \micron{} and FORCAST at 25 and 37 \micron{}), reveals that the bulk of the MIR emission is located towards the western half of the elliptical distribution observed at 8 \micron{}. There can either be more dust content towards the east of the \hii{} region, or alternatively, the UV-emitting regions can be more distributed along the Galactic plane towards the east than the heated dust in the \hii{} region. Within the Sgr C \hii{} region, three bright, warm dust clumps are especially noticeable in the ISOGAL image at 15 \micron{} in the second row of Figure~\ref{fig:sgr_allfields_HII}, and they are also perceptible as more diffuse features in FORCAST images. These sources become ill-defined in the 70 \micron{} {\it Herschel}/PACS image. In the FIR emission from relatively low-temperature dust, the entirety of the Sgr C \hii{} region becomes a silhouette except for the G359.44-0.10 EGO and its cometary-like cloud, until re-emerging as free-free and synchrotron emission at wavelengths of 2 mm and longer. In the radio continuum, there are several quasi-filamentary structures concentrated in the northwestern portion of the \hii{} region.

\begin{figure*}[t!]
    \centering
    \includegraphics[width=\textwidth]{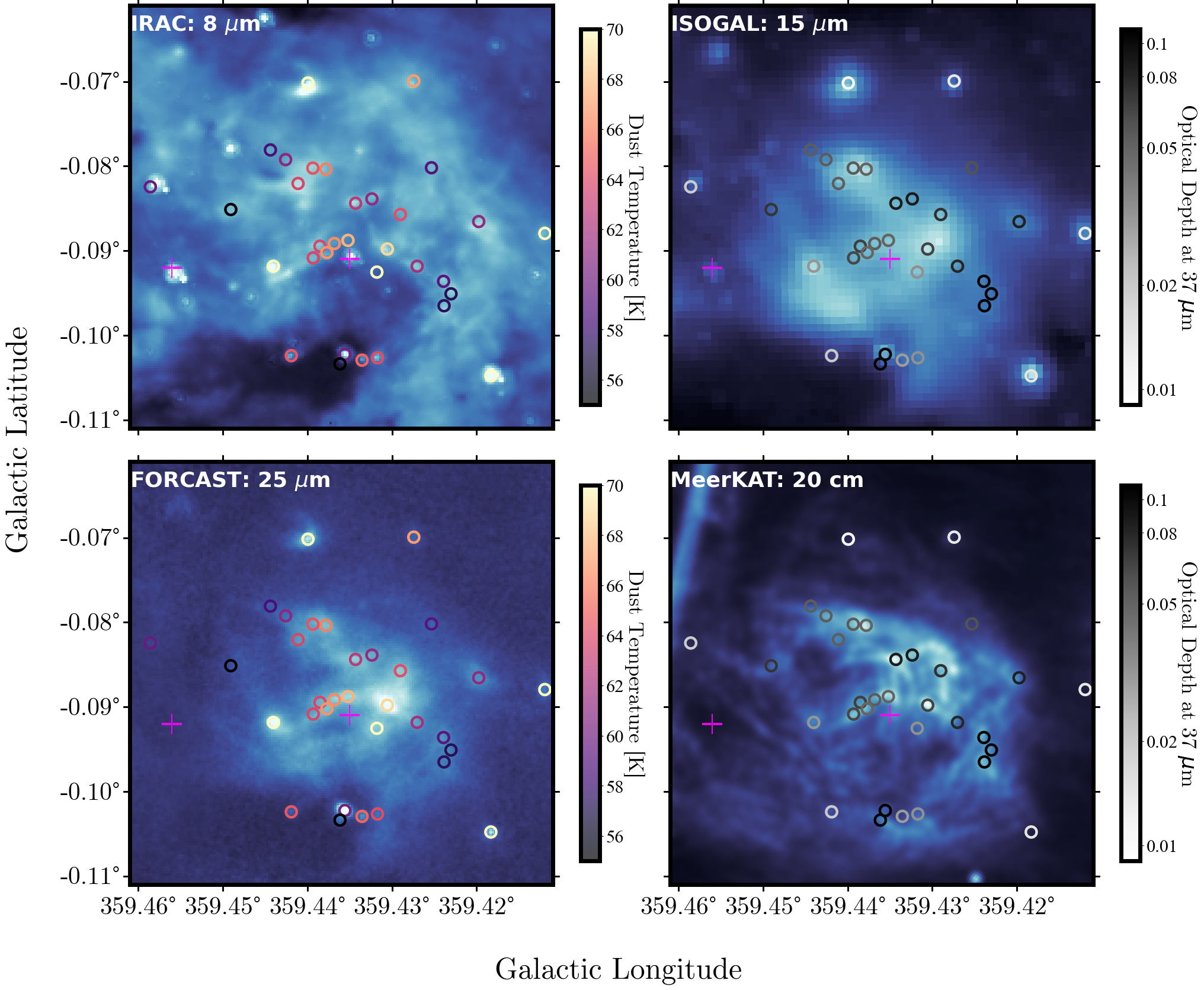}
    \caption{Selected images of the Sgr C \hii{} region at 8 \micron{} \citep{Stolovy2006}, 15 \micron{} \citep{Omont2003, Schuller2006}, 25 \micron{} \citep{Hankins2020, Cotera&Hankins2024}, and 20 cm \citep{Heywood2022}, overlaid with FORCAST-identified individual sources. The colours of the individual sources are assigned according to the dust temperature (if in colour) or optical depth (if in grey-scale) sampled at their sky-plane location, with colour bars to the right of every panel. The purple crosses show the locations of the WCL stars \citep{Geballe2019, Clark2021}. The FORCAST and MeerKAT images are displayed with a linear stretch to better visualize the density ridges within the \hii{} region. }
    \label{fig:FORCASTpt}
\end{figure*}

We next examine the structures evident in multi-wavelength detail, while overlaying the individual FORCAST sources in Figure~\ref{fig:FORCASTpt}. 
The previously mentioned dust ridge features again stand out in all MIR images, traced by an array of FORCAST sources. The FORCAST sources along the ridges have an average temperature of $65$ K and an average 37 \micron{} optical depth of 0.03. Compared to the individual sources with 8 \micron{} counterparts (e.g., stellar sources on the periphery of the \hii{} region), sources in the ridges have lower temperature and higher optical depth, supporting our previous conclusion that the individual FORCAST sources are apparently local dust density enhancements distributed along the ridges. 

Using the locations of the FORCAST sources as a reference, we find that the radio quasi-filamentary structures (lower right panel of Figure~\ref{fig:FORCASTpt}) are indeed co-spatial with the MIR dust ridges, indicating that the MIR and radio emission are from the same underlying structure. In the MeerKAT 1.28 GHz spectral index map (reported by \citealt{Heywood2022} and shown in Figure 1 in \citealt{Bally2024}), we observe that the X-shaped density ridges exhibit a higher spectral index ($\sim-0.2$) relative to the rest of the \hii{} region. \citet{Bally2024} calculated the spectral index of some of the ridges in the \hii{} region using the 1.28 GHz MeerKAT data and 98 GHz ALMA data, reported in their Table 3.
With background subtraction, these ridges tend to have a negative radio spectral index, noted by \citet{Bally2024} to signify the existence of a nonthermal component. 

We theorize that the formation of the observed dust ridges with nonthermal emission can be attributed to the central WCL star \citep{Geballe2019, Clark2021}. This star is located immediately south of the density ridge distributed along $b=-0.09$\degree{}, clearly observed in Figure~\ref{fig:FORCASTpt}. To the south of this WCL star and this emission ridge, there appears to be a cavity where the surface brightness across all MIR and radio wavelengths becomes fainter, especially at 8 \micron{} and in the radio continuum. We suggest that the stellar wind of the WCL star could be responsible for compressing and forming the dense ridge to the north and for creating the cavity to the south, where the wind is relatively unimpeded. Moreover, if the density ridges are the impact front where the high-velocity wind from the central WCL star hits and compresses the ISM, it is possible that the local electrons can be accelerated to relativistic velocities in the shocked ridges via diffusive shock acceleration \citep[e.g.,][]{Malkov2001}. Such electrons could then contribute synchrotron emission to the existing thermal emission from the ridges.  

To investigate whether there might be direct interactions between these density ridges, we have checked whether any of the intersection sites have higher radio brightness than the sum of the brightnesses of the overlapping density ridges. However, none of the intersection sites showed such an excess, so we conclude that these ridges might be joined as part of a connected system of roughly constant column density.

\subsubsection{[\cii{}] Line Analysis}
A recent study of the 158 \micron{} [\cii{}] line in the Sgr C region by \citet{Riquelme2026}\footnote{{This study was conducted using the upgraded German Receiver for Astronomy at Terahertz Frequencies (upGREAT; \citealt{Risacher2018}) aboard the SOFIA telescope, aiming to map the [\cii] fine-structure spectral line at 157.74 \micron{}. This survey has a main beam FWHM of 14\arcsec.1 and an RMS pointing accuracy of 2\arcsec.}} has revealed that the [\cii{}] emission arises predominantly from a shell around the \hii{} region across a wide range of velocities from $-60$ to $-20$ \kms{}, which is said to trace the photo-dissociation region (PDR). 

\begin{figure}
    \centering
    \includegraphics[width=\columnwidth]{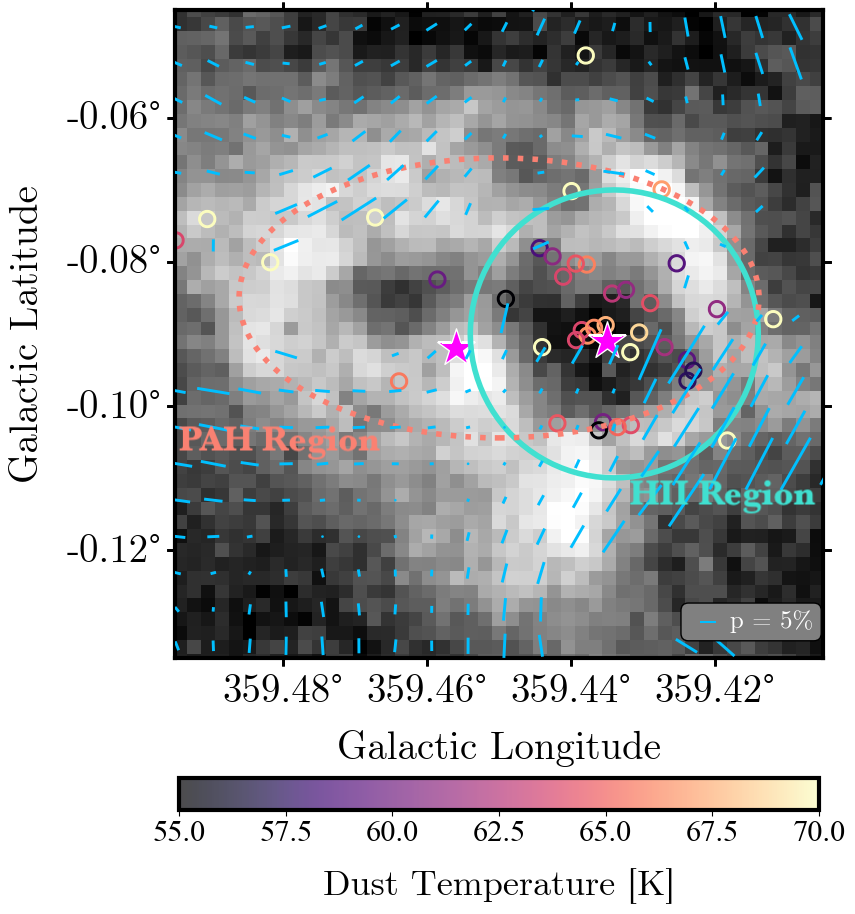}
    \caption{Greyscale image of [\cii{}] line emission in the $-41$ \kms channel, taken using the SOFIA/upGREAT instrument by \citet{Riquelme2026} overlaid with FORCAST-identified individual sources from \citet[][as circles]{Cotera&Hankins2024} and the 214 \micron{} magnetic field measurements reported by \citet[][as blue line segments]{Pare2024} {from dust emission polarimetry}, scaled by their polarization fraction. A 5\% polarized vector is plotted for reference at lower right. The FORCAST sources are coloured according to the dust temperature inferred by our FORCAST data at their location. Superimposed as magenta stars are the two WCL stars \citep{Geballe2019, Clark2021}. Additionally, we indicate the approximate extents of 8 \micron{} emission from the {\it Spitzer}/IRAC image (orange dashed ellipse labelled as ``PAH region") and emission in the MIR and radio continuum (cyan circle labelled as ``\hii{} region"). The cyan circle is also the aperture for the total IR luminosity calculation detailed in Section~\ref{sec:totlum}.}
\label{fig:CII}
\end{figure}

In Figure~\ref{fig:CII}, we compare their representative $-41$ \kms{} velocity channel image, in which the shell and its inner cavity are the most distinct, to the locations of individual FORCAST point sources and the previously mentioned WCL stars.
The [\cii{}] emission displays an overall morphology in the sky-plane that is consistent with a quasi-spherical shell. In the $-41$ \kms{} channel, there are two smaller, also quasi-spherical [\cii] cavities within this sphere, {as well as another, smaller apparent cavity to the north, located at $l=359.445$\degree{}, $b=-0.066$\degree{}}. The combined outer boundary of these cavities roughly corresponds to the elliptical emission observed in the {\it Spitzer}/IRAC image at 8 \micron{}, largely tracing PAH emission (dashed orange ellipse in Figure~\ref{fig:CII}). The majority of FORCAST sources and the central WCL star, however, are located within the larger cavity towards the west. The outer boundary of the \hii{} region seen in the MIR and radio continuum (i.e., the ionization front) is coincident radially with the brightest portion of the [\cii{}] shell (cyan circle in Figure~\ref{fig:CII}), such that the \hii{} region is largely contained within this shell. The co-spatial MIR and [\cii{}] emissions confirm that the central UV sources are heating and ionizing, or photo-dissociating the gas at and beyond this interface. Again, we note that the WCL star situated in the centre of the cavity is undoubtedly an important contributor to the UV flux. 

At the northwestern portion of the \hii{} region along $l\sim0.08$\degree{}, we observe an array of FORCAST sources immediately bordering the boundary of the PDR traced by [\cii{}] emission, where this portion of the PDR also overlaps with the PAH, MIR, and radio emission on the boundary of the \hii{} region. As these FORCAST sources predominantly trace the density ridges, we postulate that the border between the density ridges and the PDR signifies a compression front driven by the combined actions of the ionization front and the stellar wind(s). {\citet{Riquelme2026} and \citet{Zhu2026} both consider the hypothesis of a possible prior supernova. The former invokes the supernova hypothesis to account for the momentum of the expanding shell, and the latter proposes the same conclusion drawn from an X-ray analysis.} \cite{Zhao2025} found that the magnetic field is largely tangential to the [\cii{}] shell, as well as to one of the dust ridges on the border of the \hii{} region. {They also inferred a plane-of-sky magnetic field strength of $\sim30–400~\mu{\rm G}$ in the surrounding molecular material. In the relativistic limit, the associated Larmor radii at the MeerKAT central frequency of 1.28 GHz is therefore orders of magnitude smaller than the radius of the \hii{} region. The electrons should therefore be strongly tied to the local field lines, and the tangential field may act as a magnetic barrier to cross-field escape.} If the spectral index values for the radio emission from the ridges can be further confirmed to be as steep as was indicated by \citet{Heywood2022}, it may suggest that the magnetic field identified in the molecular cloud confines the relativistic electrons inside the \hii{} region and its surrounding PDR.  

The smaller [\cii{}] cavity to the east is devoid of any clearly associated MIR or radio sources. However, together with the large western cavity, their combined spatial extent matches the extent of the PAH region (dashed orange ellipse; Figure~\ref{fig:CII}). The ensemble of observations suggests that the smaller eastern cavity may consist of interstellar material that is illuminated by longer-wavelength UV, either from the eastern WCL star not within the \hii{} region (2-MASS label J17444083-2926550), or by photons from the central WCL star that have penetrated beyond the eastern shell of the \hii{} region.

\subsubsection{Total IR Luminosity}
\label{sec:totlum}
To constrain the stellar population residing inside the Sgr C \hii{} region, we perform an SED analysis using the FORCAST data together with other IR surveys, to estimate the total IR luminosity of the \hii{} region and to thereby constrain the stellar population in this region.

We use a wide range of images, including those from {\it Spitzer}/IRAC at 3.6, 4.5, 5.8, and 8 \micron{} \citep{Stolovy2006}, the ISOGAL survey at 15 \micron{} \citep{Omont2003, Schuller2006}, SOFIA/FORCAST at 25 and 37 \micron{} \citep{Hankins2020, Cotera&Hankins2024}, and the {\it Herschel}/PACS Hi-GAL survey at 70 and 160 \micron{} \citep{Molinari2010, Molinari2016}. To correct for dust extinction at the short-wavelength end, we again apply the interpolated $\tau_{\rm sil}$ extinction map {described in Section~\ref{sec:extinction}} to images with wavelengths up to 37 \micron{}. Following the extinction law of \citet{Chiar2006}, we adopt $\tau_8/\tau_{\rm sil}=0.2969, \tau_{15}/\tau_{\rm sil}=0.336$ as conversion factors for {\it Spitzer}/IRAC and ISOGAL images\footnote{ which are also verified using the extinction curve of \citet{Fritz2011}}, respectively. The conversion factors for FORCAST wavelengths are stated in Section~\ref{sec:colourtemp}. Given that extinction at 70 \micron{} is smaller than at the shorter wavelengths, we do not attempt to apply an extinction correction to the 70 and 160 \micron{} images. To ensure a uniform image resolution, we convolve all included images to match the lowest resolution of the Hi-GAL 160 \micron{} image, 12\arcsec{}. 

For this analysis, we perform the photometry computation using the \texttt{Photutils} package \citep{Photutils} with an aperture centred at $l=359.434$ \degree{}, $b=-0.090$ with a 0.21\degree{} radius {to cover the \hii{} region with about 10 pc in diameter} (cyan circle in Figure~\ref{fig:CII}). Given the unique and complex environment of Sgr C, we perform careful background subtraction at each wavelength using standard photometry techniques. Namely, we select an annulus that is relatively noise-dominated compared to the \hii{} region, calculate the median flux per pixel within that annulus, and then subtract that value from each pixel within the central aperture of the \hii{} region. We use an annulus with a 0.3\degree{} outer radius and a 0.25\degree{} inner radius. {However, to adjust for the instrumental properties of the FORCAST instrument (i.e., variable noise properties across frames and over-subtracted background induced by the aforementioned ``bowl" effect; see Section~\ref{sec:obs}), for FORCAST images only, we use a smaller annulus with 0.25\degree{} outer radius and a 0.21\degree{} inner radius to avoid contamination from these effects.} Additionally, for the {\it Spitzer}/IRAC 3.6 and 4.5 images, in order to avoid contamination from the Rayleigh-Jeans emission from the abundant stellar sources, we replace pixels that are brighter than a $3\sigma$ limit from the mean with the median flux density value within the aperture. 

We model the SED of the extracted fluxes using the \texttt{DustEM} software \citep{dustEM}. {\texttt{DustEM} is equipped with the capability to model very small transiently-heated grains, including PAHs, which allows for the most inclusive luminosity estimation at wavelengths shorter than 8 \micron{}.} As defined by \citet{Compiegne2011}, this model includes four grain species. We list them in terms of their default mass fractions from the package\footnote{The mass fraction of a species $A$, $Y_A=M_A/M_H$, is defined relative to the mass of hydrogen, $M_H$.}: PAH ($Y_\text{PAH}=7.8\times10^{-4}$), small amorphous carbon (SamC; $Y_\text{SamC}=1.65\times10^{-4}$), large amorphous carbon (LamC; $Y_\text{Lamc}=1.45\times10^{-4}$), and astronomical silicates (aSil; $Y_\text{aSil}=7.8\times10^{-3}$). We also model the interstellar radiation field ($G_0$; as defined by \citealt{Mathis1983}) scaled with a normalization factor $n$. We fit our SED model to observations by multiplying their mass fractions with a free scaling factor denoted $X_A$, where $A$ is the species, as the parameters to be fit. However, out of the four species, the mass fractions of LamC and aSil are adopted directly from \citet{Compiegne2011} as listed above for the diffuse ISM without rescaling, since the emissions from these species are largely indistinguishable in shape, so the degeneracy between these two species is difficult to break with photometry alone. To summarize, there are four free parameters to be fit to the extracted flux: the interstellar radiation field $G_0$, the radiation field normalization $n$, and the scaling factors of PAH and SamC ($X_\text{PAH},\, X_\text{SamC}$). 

\begin{figure}
    \centering
    \includegraphics[width=\columnwidth]{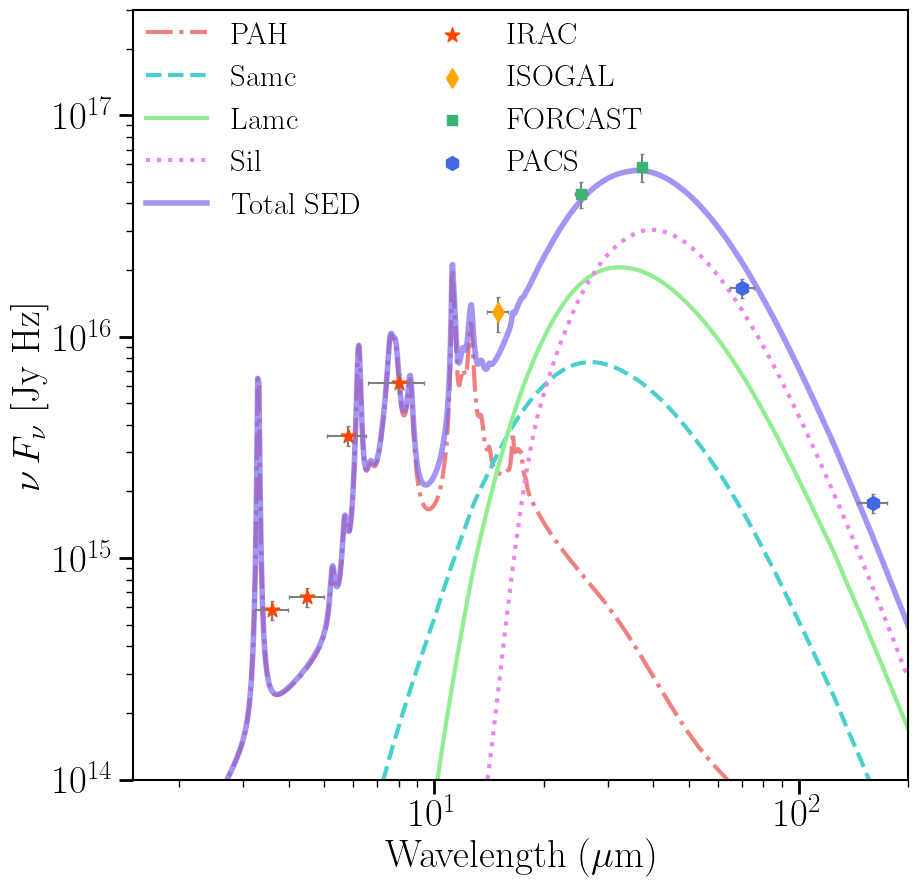}
    \caption{The SED of the Sgr C \hii{} region fitted by \texttt{DustEM}. The surveys from which each point is extracted can be found in the legend at the top left. The thickened purple curve is the total SED, and the contributions of the various grain types are indicated by the colours shown also in the legend. }
    \label{fig:dustEM}
\end{figure}

To find the best-fit parameter values, we use the tree-structured Parzen Estimator optimization algorithm implemented by the \texttt{Optuna} optimizer \citep{optuna}  to minimize the {reduced} $\chi^2$ value between the modelled SED and the collected photometric data. After 300 iterations, we find that the following grain abundances and parameter values provide the best-fit for the extracted fluxes: PAH scaling factor ($X_\text{PAH} = 0.58$), SamC grains scaling factor ($X_\text{SamC} = 2$), the radiation field parameter ($G_0=1.5\times10^3$), and normalization factor ($n=1.5$). We present the best-fit SED in Figure~\ref{fig:dustEM} with a $\chi^2$ value of 8.63.
The grain abundances are in accordance with those found in other, smaller \hii{} regions observed in the MIR near Sgr A and the Arches cluster \citep{Hankins2017, Hankins2019}. Here, we note that the {\it Spitzer}/IRAC 3.6 and 4.5 \micron{} data points are higher than the best-fit SED, possibly suggesting persisting stellar contamination. If these two points are excluded from the $\chi^2$ calculation, we obtain a reduced $\chi^2$ value of 5.1. We find the \hii{} region to have a total IR luminosity of $(1.40\pm0.19)\times10^{6}L_\odot$ and a dust mass of $5.4\pm 0.7 M_\odot$. 

Previous studies of the Wolf-Rayet star population in the GC found that the luminosity of {single} WC stars typically ranges from $10^{4.9}$ to $10^{5.6} L_\odot$ \citep{Crowther2007, Sander2012}. From their reported results, we conclude that the central WCL star is unlikely to be the only stellar source illuminating the \hii{} region, as we are unaware of any WCL star with luminosity exceeding $10^{6}L_\odot$. Therefore, there likely exists undetected relatively massive young stars within the \hii{} region accounting for the high total IR luminosity.

\subsection{EGO: G359.44-0.10 }
Another Sgr C feature that is of particular interest is the G359.44-0.10 EGO, which is observed at almost every wavelength except in the radio continuum (see Figure~\ref{fig:sgr_allfields_HII}). In the NIR and MIR, the EGO stands out sharply as a bright and extremely compact source against a cloud silhouette extending from the EGO towards the southeast. This absorbing structure is commonly referred to as the Sgr C molecular cloud, known for its cometary appearance, high cloud density, and FIR luminosity \citep{Zhao2025}. Recent surveys have reported that the EGO consists of two cores that are resolved by both JWST-NIRCam \citep{Crowe2024} and SCUBA (fourth row of Figure~\ref{fig:sgr_allfields_HII}). \citet{Crowe2024} also reported comprehensive SED fitting results for the EGO, {including the FIR wavelengths}\footnote{ In this work, we do not attempt to perform an SED fit of the EGO and other sources due to the complexity of performing an adequate background subtraction, given both the extended background emission and instrumental effects (e.g., non-uniform resolution across surveys and the aforementioned ``bowl" effect of the FORCAST camera). }. For each core, they estimate a stellar mass of approximately $20\, M_\odot$ and a bolometric luminosity of $\sim10^5L_\odot$. Our FORCAST dust temperature analysis  (Figure~\ref{fig:combined_T_tau}) indicates that the EGO has a lower temperature, $\lesssim$50 K, and a higher optical depth, $\gtrsim$0.05, than any other region within Sgr C. This is consistent with the current consensus that this EGO harbours massive protostars \citep{Kendrew2013}. 

\subsection{YSO Cluster: G359.43+0.02}
\label{sec:YSO}
The SOFIA/FORCAST observations carried out in Cycle 9 covered the G359.43+0.02 YSO cluster. On the sky-plane, the northern extension of the Sgr C NTF passes through this region with signs of possible interaction reported by \citet{Zhao2025}, and discussed below. However, as previously mentioned, the relatively low sensitivity of the FORCAST instrument leaves its images with lower SNR in this region compared to the \hii{} region. Therefore, the derived dust temperatures and optical depths (see Figure~\ref{fig:combined_T_tau}) are only significant for the brightest cluster sources.

\begin{figure*}
    \centering
    \includegraphics[width=\textwidth]{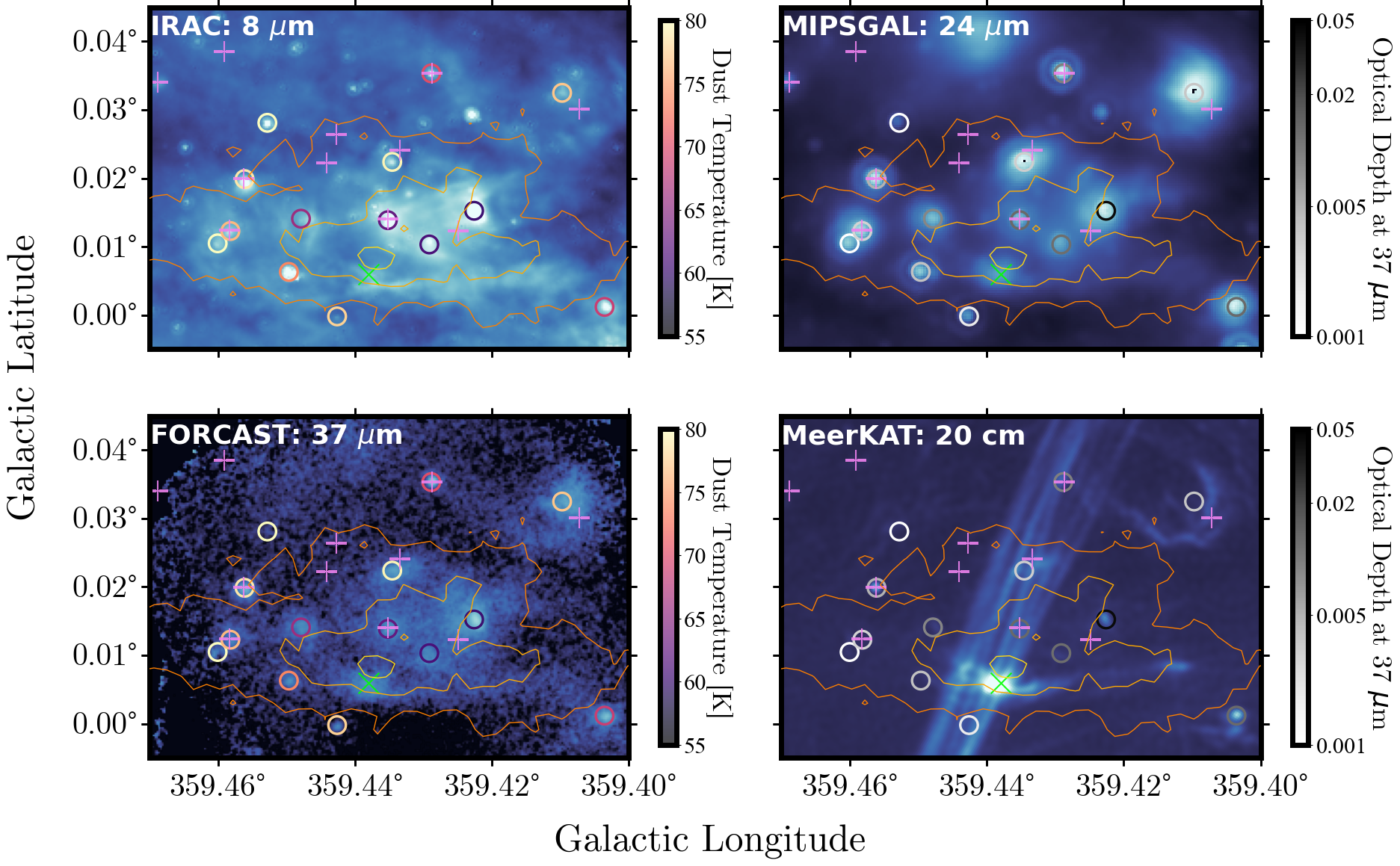}
    \caption{Selected images of the G359.43+0.02 YSO cluster at 8 \micron{} \citep{Stolovy2006}, 24 \micron{} \citep{Carey2009}, 37 \micron{} \citep{Hankins2020}, and 20 cm \citep{Heywood2022}, overlaid with FORCAST-identified individual sources. {The contours are the velocity-integrated [\cii] emission from $-100$ to $-80$ \kms{}, drawn at (30, 82, and 133) K \kms{} from red to yellow.} The coloured circles denote the dust temperature, and the grey-scale circles give the optical depth, as indicated by the scale bars. The pink crosses mark the locations of the YSOs reported by \cite{Yusef-Zadeh2009}, based on the SED fitting results of both {\it Spitzer}/IRAC and {\it Spitzer}/MIPS observations from 3.6 to 24 \micron{}. The green cross marks the source G359.438+0.006 (or FIR-4). }
    \label{fig:YSO}
\end{figure*}

Figure~\ref{fig:YSO} shows multi-wavelength images of the G359.43+0.02 cluster overlaid with markers for the individual FORCAST sources. 
This region is filled with diffuse PAH emission observed in the 8 \micron{} image, which shows several linear ridge-like structures. The 24 and 37 \micron{} emission features tend to be concentrated around the FORCAST-identified sources, likely tracing heated circumstellar dust. The radio emission here is dominated by the prominent NTF and the horizontal filament along $l=0.01$\degree{} associated with G359.438+0.006, or FIR-4. There are some compact radio sources noted by \cite{Yusef-Zadeh2009} that have MIR counterparts. In general, the sources in this cluster typically exhibit a much higher temperature ($\gtrsim70$ K) and lower optical depth ($\sim10^{-3}$) than the \hii{} region. In particular, the four FORCAST sources along $l=0.015$\degree{} are the only ones having temperature lower than 65 K (and correspondingly higher optical depths), while the rest have colour temperatures higher than 75 K. We compare the locations of FORCAST sources with those classified as YSOs by \citet{Yusef-Zadeh2009} and labelled with pink crosses in Figure~\ref{fig:YSO}. All YSO sources that are also catalogued as FORCAST sources are classified as young stage I objects by \citet{Yusef-Zadeh2009}.

{We note the substantial presence of [\cii] emission in this region, as plotted in Figure~\ref{fig:YSO} as contours. The integration over [\cii{}] velocity ranges from $-100$ to $-80$ \kms{}, and is selected to match the velocity of this region reported by \cite{Riquelme2026}. We find that the [\cii] emission generally has a shape similar to that of the MIR emission, presumably as a dust ridge coincident with an ionization front formed by the winds and UV radiation from the YSO cluster. The UV radiation from this star cluster is the likely cause of both the [\cii{}] emission and dust heating, creating an extended photodissociation region. }

One particular source of interest is G359.438+0.006. This source is located on the eastern end of a MIR ``stripe" that extends from $l=359.41~\mathrm{to}~359.44$\degree{} along $b=0.005$\degree{}. In the radio continuum, this source is located at the intersection of the radio counterpart of the MIR ``stripe" and the Sgr C NTF. However, due to the low SNR of the FORCAST data in this region, although it is catalogued as a FORCAST source by \citet{Cotera&Hankins2024}, the emission does not pass our additional convolution and 3$\sigma$ cut due to its low luminosity and SNR. \citet{Zhao2025} found that the radio surface brightness at the location of the source {is significantly higher than the summed intensities of the two composite filaments -- the nonthermal Sgr C NTF and the thermal G359.438+0.006, strongly suggesting a physical interaction between them. Additionally, we note that the peak [\cii{}] emission lies immediately north of the the radio peak G359.438+0.006 (green cross in Figure~\ref{fig:YSO}).  The radio emission associated with G359.438+0.006 extends slightly northward along the NTF into the small elliptical [\cii{}] contour surrounding the peak, raising the possibility that electrons generated by UV ionization at this location might be accelerated to relativistic energies by magnetic field line reconnection, and fed into the bundle of radio filaments at this location to produce the synchrotron emission that illuminates the NTF, in a manner analogous to that proposed for the Sickle \hii{} region, G0.18-0.04 \citep{SerabynMorris94}. We also observe elevated MIR luminosity at the interaction site, which can perhaps ascribable to the interaction.}

\subsection{Source C: G359.38-0.07}
\begin{figure*}
    \centering
    \includegraphics[width=\textwidth]{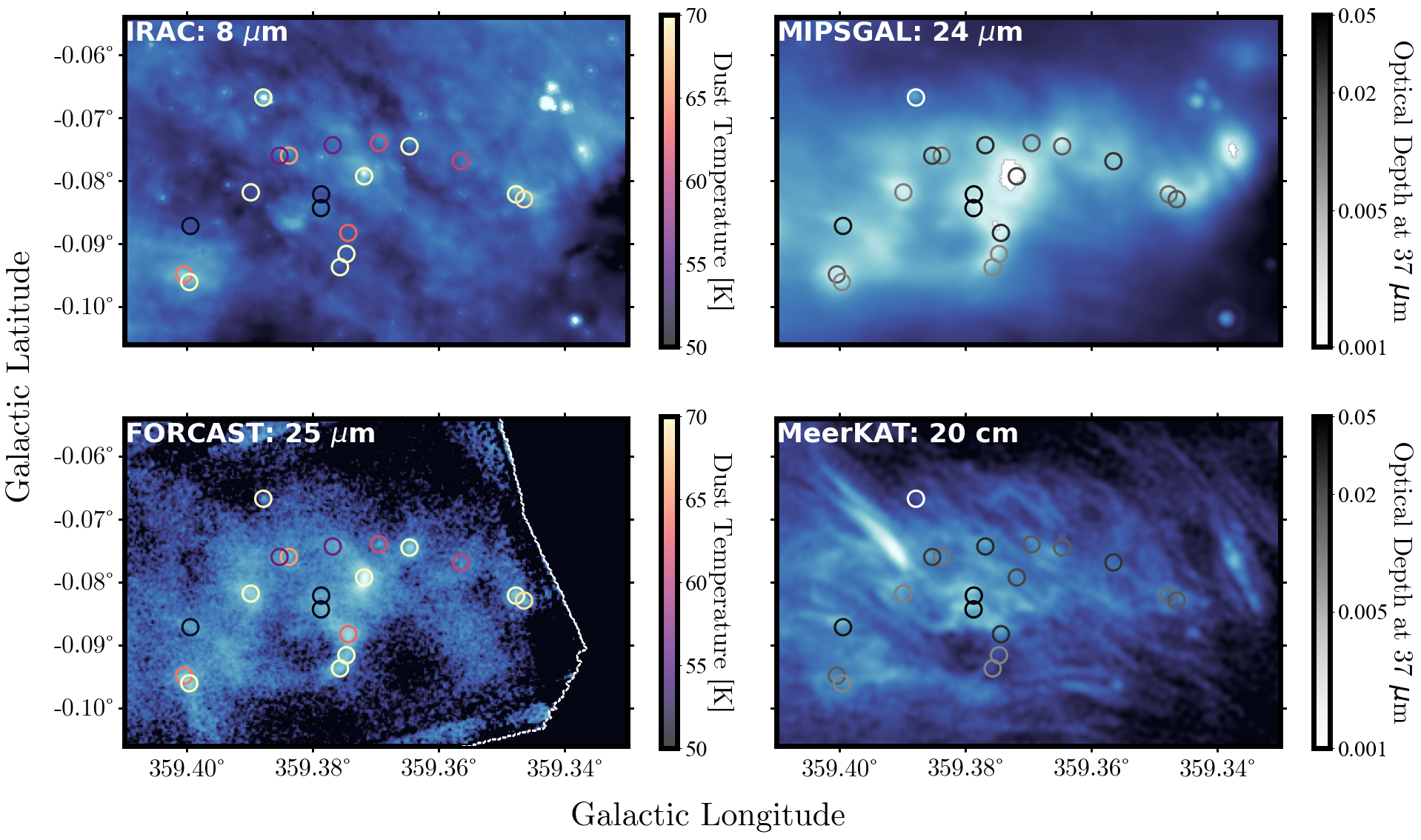}
    \caption{Selected images of Source C {and the adjacent region} marked  at 8 \micron{} \citep{Stolovy2006}, 24 \micron{} \citep{Carey2009}, 25 \micron{} \citep{Hankins2020}, and 20 cm \citep{Heywood2022}, overlaid with FORCAST-identified individual sources. The circle colours, or shades, carry the same meaning as in Figure~\ref{fig:YSO}}
    \label{fig:sourceC}
\end{figure*}

Toward the immediate east of the Sgr C \hii{} region is a warm thermal MIR and radio source denoted Source C \citep{Roy2003, Lang2010}, confirmed by \citet{Lang2010} to have a line-of-sight distance similar to that of the \hii{} region. Much like the \hii{} region, Source C is absent in the FIR. Figure~\ref{fig:sourceC} shows multi-wavelength images of Source C overlaid with circles showing the locations of individual FORCAST sources, from \citet{Cotera&Hankins2024}. 

The general MIR morphology of Source C consists of a bright ``head" at $l=359.37$\degree{}, $b=-0.08$\degree{} and a few diffuse ``tails" extending eastward. The ``head" has a compact counterpart in the {\it Spitzer}/IRAC image, likely a stellar source responsible for the surrounding ionization. The temperature of the dust surrounding this ``head" reaches $70$ K. The optical depths of the sources inside this warm cloud are all of the same order of magnitude, ranging between $0.02$ and $0.05$. In the radio continuum, Source C is comprised of a bundle of quasi-parallel striations oriented northeast-southwest. Source C is also associated with a luminous, short radio filament (lower right panel of Figure~\ref{fig:sourceC}). However, we do not observe any MIR features that could be associated with the radio striations, especially in the {\it Spitzer}/IRAC bands. Based on our multi-wavelength observation, we suggest that Source C is likely a relatively low-density region containing gas that has been ionized by a local source, or sources, possibly the compact source in the ``head". 

\section{Conclusions}
\label{sec:conclusion}
This paper presents an analysis of MIR-bright regions in the Sgr C complex using the SOFIA/FORCAST Galactic Center Legacy Survey data product in combination with imaging data from other observatories. Taking advantage of the absence of bright source saturation with FORCAST, this study focuses largely on the MIR-bright Sgr C \hii{} region, while covering other adjacent MIR-bright sources such as the G359.43+0.02 YSO cluster and Source C. We summarize our findings as the following:
\begin{itemize}
    \item We find that dust in the MIR bright regions has a temperature range of $50-90$ K. The Sgr C \hii{} region has a peak temperature of approximately $70$ K near the central WCL star, and gradually decreases to $\sim$55 K towards its periphery, especially towards the west. We calculate a heating profile of a late-type, carbon-sequence Wolf-Rayet star (denoted the central WCL star in this work) located in the centre of the Sgr C \hii{} region in projection, and find that the heating profile agrees well with the computed dust temperature map assuming smaller grain size ($\sim$0.01\micron{}), suggesting this star could be a main contributor of heating in the Sgr C \hii{} region. 
    \item The typical 37 \micron{} optical depth of the Sgr C \hii{} region is around $10^{-2}$. We find several density ridges that have an elevated optical depth of $\sim10^{-1}$ in the central portion of the \hii{} region. Additionally, the western portion of the \hii{} region exhibits higher optical depth than the central and eastern portions. 
    \item We find multiple dust emission ridges inside the Sgr C \hii{} region with counterparts in the radio continuum, which is also reported by \cite{Bally2024} using JWST-NIRCam images. The dust ridges have an elevated optical depth compared to their surrounding area, but do not exhibit significant temperature variation from their surroundings. We postulate that these dust ridges are density ridges of the ionized gas and embedded dust enhanced by the stellar winds within the region. Many compact and extended MIR sources are found as local flux peaks along these ridges. We suggest that the wind from the central WCL star may accelerate the free electrons in the ridges via diffusive shock, creating nonthermal emission in the ridges as suggested by \cite{Bally2024}. 
    \item On the periphery of the Sgr C \hii{} region, we find that one of the densest dust ridges immediately borders the PDR signified by a [\cii{}] shell tracing the periphery of the Sgr C \hii{} region \citep{Riquelme2026}. Together with the observation that there is a magnetic field largely tangential to the surface of the \hii{} region \citep{Zhao2025}, we postulate that such a magnetic field can confine the non-thermal synchrotron radiation within the boundary of the \hii{} region, subject to further confirmation of the radio spectral index. 
    \item From fitting a \texttt{DustEM} SED model, we find the total MIR luminosity of the Sgr C \hii{} region to be $(1.40\pm0.19)\times10^{6} L_\odot$ and the dust mass to be $5.4\pm 0.7 M_\odot$. The central WCL star, albeit contributing substantially to the high MIR luminosity and heating of the region, is unlikely to be the sole provider of a luminosity of this magnitude. This inference implies the probable presence of additional massive stars within the \hii{} region. 
    \item We survey other intriguing MIR bright sources in the Sgr C region. The G359.44-0.10 EGO has the lowest temperature ($\lesssim50$ K) and highest optical depth ($\gtrsim0.5$), confirming previous reports that they are dense clouds that favour star formation \citep{Kendrew2013, Crowe2024}. The sources found in the G359.43+0.02 YSO cluster, on the periphery of the \hii{} region, and along the Galactic plane are mostly stellar sources with {\it Spitzer}/IRAC counterparts. They have higher temperatures and lower optical depths than the \hii{} region. As for Source C, we suggest that it is likely a low-density region of ionized gas, possibly due to the compact source in the high temperature ``head" having a {\it Spitzer}/IRAC counterpart source.
\end{itemize}

Past and ongoing star formation is clearly central to the phenomenology of the Sgr C \hii{} region and its surroundings, but a few important parts of the story remain to be elucidated. Apart from the two known WCL stars in the region, we can infer that more -- so far unidentified -- massive stars remain to be found to account for the total luminosity. The suggestion by \citet{Riquelme2026} that it might be necessary to invoke a past supernova to account for the expansion of the [\cii{}] shell around the \hii{} region could reveal the fate of at least one massive star.  

Finally, the massive protostellar objects in the EGO herald the next phase in the evolution of this key region lying at the interface between an expanding \hii{} region and a molecular cloud. { On larger scales, the ongoing activity of the Sgr C complex has presumably been induced, at least in part, by the shocks resulting from the impact of the Galactic-bar-induced inflow onto the CMZ \citep[e.g.,][]{Tress2020}, in which case the phenomena investigated here are part of a continuing sequence of star-formation events.}

\begin{acknowledgments}
This work is primarily based on observations made with the NASA/DLR Stratospheric Observatory for Infrared Astronomy (SOFIA). SOFIA was jointly operated by the Universities Space Research Association, Inc. (USRA), under NASA contract NNA17BF53C, and the Deutsches SOFIA Institut (DSI) under DLR contract 50 OK 2002 to the University of Stuttgart. Financial support for this work was provided by NASA contract NNA17BF53C through SOFIA Awards \#07$\_$0189 and \#09$\_$0216 issued by USRA. This work has made extensive use of NASA's Astrophysics Data System (\href{http://ui.adsabs.harvard.edu/}{http://ui.adsabs.harvard.edu/}) and the arXiv e-Print service (\href{http://arxiv.org}{http://arxiv.org}).

R.J.Z. and M.R.M. were supported by NASA through award \#063021 administered as a subgrant from Arkansas Tech University (ATU) to UCLA. R.J.Z. is also supported by The University of Chicago for partial duration of this work. R.J.Z. thanks Eva Liu for acquiring supporting resources and Rory Bentley for helpful conversations. Finally, R.J.Z. wishes to acknowledge all community members of Meadowridge School, located in British Columbia, Canada, for their home-like hospitality and helpful conversations while much of this work was completed. 

The authors at The University of Chicago acknowledge our presence on the ancestral and unceded territories of the Kickapoo, Peoria, Potawatomi, Miami, and Sioux people. Much of this work was completed at Meadowridge School, located on the ancestral and unceded territories of the Katzie, Kwantlen, and Coast Salish Peoples, as well as UCLA, located on the traditional, ancestral, and unceded territory of the Gabrielino/Tongva peoples. We value the opportunity to learn, live, and share research and educational experiences on these traditional lands.
\end{acknowledgments}

\begin{contribution}
R.J.Z. performed majority of data analysis, produced all figures and tables in the paper, and drafted the manuscript. M.R.M. provided continuous supervision and guidance, led the editing process of the manuscript, acquired funding to support this project, and offered feedback on all aspects. M.J.H. carried out additional mosaic improvements, performed data processing and analysis, and along with A.S.C. guided key analysis decisions on using SOFIA/FORCAST data products. J.P.S. provided the interpolated version of the dust extinction map. All authors contributed to the data interpretation and editing process of this manuscript. 
\end{contribution}

\facility{SOFIA, APEX, {\it CXO}, {\it  Herschel}, IRAM, {\it ISO}, JCMT, MeerKAT, Mopra, {\it Spitzer}, VLA}

\software{\texttt{Astropy} \citep{Astropy2013, Astropy2018}, \texttt{cmocean} \citep{cmocean}, \texttt{DustEM} \citep{Compiegne2011}, \texttt{Matplotlib} \citep{Matplotlib}, \texttt{Numpy} \citep{Numpy}, \texttt{Optuna} \citep{optuna}, \texttt{Pandas} \citep{Pandas}, \texttt{Redux} \citep{Clarke2015}, \texttt{Photutils} \citep{Photutils}, \texttt{Reproject} \citep{Reproject}, \texttt{Scipy} \citep{SciPy}}

\vspace{5mm}
\counterwithin{figure}{section}
\counterwithin{table}{section}

\renewcommand{\thesection}{A.\arabic{section}}
\renewcommand{\thefigure}{A.\arabic{figure}}
\renewcommand{\thetable}{A.\arabic{table}}
\addtocounter{table}{-1}

\bibliography{refs}{}
\bibliographystyle{aasjournal}
\end{document}